\documentclass[usenatbib]{emulateapj}

\slugcomment{{\sc Accepted to ApJ:} January 28, 2016}

\shorttitle{Draft}
\shortauthors{A. R. Hill et al.}
\begin{document}
\title{A Stellar Velocity Dispersion for a Strongly-Lensed, Intermediate-Mass Quiescent Galaxy at $\lowercase{z}=2.8$}

\author{Allison. R. Hill$^{1}$, Adam Muzzin$^{2}$, Marijn Franx$^{1}$, Jesse van de Sande$^{3}$}
\affil{$^{1}$Leiden Observatory, Leiden University, P.O. Box 9513, 2300 RA,
Leiden, The Netherlands}
\affil{$^{2}$Institute of Astronomy, University of Cambridge, Madingley Road, CB3 0HA,
Cambridge, UK}
\affil{$^{3}$Sydney Institute for Astronomy, School of Physics, University of Sydney, NSW 2006,
Australia
}
\email{hill@strw.leidenuniv.nl}

\begin{abstract}

Measuring stellar velocity dispersions of quiescent galaxies beyond $z\sim2$ is observationally challenging. Such measurements require near-infrared spectra with a continuum detection of at least moderate signal-to-noise, often necessitating long integrations. In this paper, we present deep X-Shooter spectroscopy of one of only two known gravitationally-lensed massive quiescent galaxies at $z>2$. This galaxy is quadruply imaged, with the brightest images magnified by a factor of $\sim5$. The total exposure time of our data is 9.8 hours on-source; however the magnification, and the slit placement encompassing 2 images, provides a total equivalent exposure time of 215 hours. From this deep spectrum we measure a redshift ($z_{\mathrm{spec}}=2.756\pm0.001$), making this one of the highest redshift quiescent galaxies that is spectroscopically confirmed. We simultaneously fit both the spectroscopic and photometric data to determine stellar population parameters and conclude this galaxy is relatively young ($560^{+100}_{-80}~\mathrm{Myr}$), intermediate-mass ($\log{M_{*}/\mathrm{M_{\odot}}}=10.59^{+0.04}_{-0.05}$), consistent with low dust content ($A_{V}=0.20^{+0.26}_{-0.20}$), and has quenched only relatively recently. This recent quenching is confirmed by strong Balmer absorption, particularly $H\delta$ ($H\delta_{A}=6.66^{+0.96}_{-0.92}$). Remarkably, this proves that at least some intermediate-mass galaxies have already quenched as early as $z\sim2.8$. Additionally, we have measured a velocity dispersion ($\sigma=187\pm43~\mathrm{km/s}$), making this the highest-redshift quiescent galaxy with a dispersion measurement. We confirm that this galaxy falls on the same mass fundamental plane (MFP) as galaxies at z=2.2, consistent with little to no evolution in the MFP up to z=2.8. Overall this galaxy is proof of existence of intermediate-mass quenched galaxies in the distant universe, and that lensing is a powerful tool for determining their properties with improved accuracy.  

\end{abstract}

\keywords{galaxies: evolution, galaxies: formation, gravitational lensing: strong, galaxies: kinematics and dynamics, galaxies: stellar content, galaxies: structure}

\section{Introduction}
\label{sec:intro}

It is well established that galaxies with evolved stellar populations were already in place when the universe was less than half of its current age (see for example \citealt{mccracken2012,ilbert2013,muzzin2013b}). These galaxies were first identified from a population which exhibited red $\mathrm{J_{s}-K_{s}}$ colors \citep{franx2003}. These colors were consistent with taking the spectral energy distribution (SED) of an elliptical or dusty star-burst galaxy and red-shifting to $z\sim2$, with the degeneracy between the two types of galaxies lifted with the inclusion of IRAC data \citep{labbe2005, williams2009}. Out of the \citet{franx2003} sample, subsequent spectroscopy later confirmed that a subset of these `red-galaxies' were indeed quiescent \citep{vandokkum2003, kriek2006}, establishing that galaxies with strongly suppressed star formation were present at higher redshifts. 

A comparison of the stellar populations of quiescent galaxies between $z\sim2$ and $z\sim0$ via their mass-to-light ratios show that the stellar populations in these galaxies are consistent with passive evolution \citep[e.g.,][]{vandokkum1998, treu1999, bernardi2003, vandokkum2006}. Although quiescent, the $z\sim2$ galaxies are strikingly different in their structure compared to present-day ellipticals. At a given mass, their effective radii ($\mathrm{r_{e}}$) are a factor of $\sim2-4$ smaller than at $z\sim0$ \citep[e.g.,][]{daddi2005, trujillo2006, zirm2007, vandokkum2008, vandokkum2010, szomoru2012}. This difference in $\mathrm{r_{e}}$ between present-day ellipticals and high-redshift galaxies implies a rapid structural evolution between $z\sim2$ and today. Although these galaxies must grow by a factor of a few in size, their central stellar-velocity dispersions show little evolution \citep{toft2012, vandesande2013, bezanson2013b, belli2014a} between these redshifts. 

The evolution of stellar populations between redshift $z\sim2$ and $z\sim0$ is mirrored in the evolution of the zero-point in the fundamental plane (FP). The FP represents a locus of galaxies which occupy a tight plane determined by a galaxy's surface brightness, size and velocity dispersion \citep{dressler1987,djorgovski1987}.This plane maintains a slight tilt with respect to the expectation from the assumption of virial equilibrium. This tilt is thought to be caused by a deviation from homology \citep{pahre1995,capelato1995,busarello1997}, and by variations in the mass-to-light ratio \citep{vandokkum1998,cappellari2006,robertson2006,bolton2007, cappellari2013}.  

When the dependence of the FP on surface brightness is replaced with the average stellar mass density (i.e. the mass fundamental plane; henceforth MFP), the tilt in the FP virtually disappears \citep{bolton2007}, or is at least shown to be weaker than the tilt in the FP \citep{bolton2008,holden2010,bezanson2013b}. In contrast to the evolution in the zero-point of the FP \citep[e.g.,][]{vandesande2014}, the MFP zero-point shows very little evolution out to $z\sim2$ \citep{bezanson2013b}, reminiscent of the lack of evolution in central stellar-velocity dispersion \citep{toft2012, bezanson2013b, belli2014a}. However the precise evolution depends on assumptions when counting galaxies and how to connect progenitors to their descendants \citep{vandesande2014}. 

The consistency of the slope of the MFP with that predicted from virial equilibrium points to variations in mass-to-light ratios as the likely cause of the tilt in the FP. The evolution in mass-to-light ratios are driven by either variations in dark matter content, variations in stellar populations, or a combination of both. In the context of MFP evolution at high-z, it is important to note that the sample of \citealt{bezanson2013b} in the highest redshift bin was restricted to massive galaxies i.e $\log{M_{*}/M_{\odot}}>11.0$, leaving the MFP unpopulated below this mass threshold at $z\sim2$, and due to redshift coverage, at all masses above $z\sim2$.

Although valuable for testing the evolution of the MPF, measuring stellar velocity dispersions of quiescent galaxies beyond $z\sim2$ has proven technically challenging. At these redshifts their optical absorption lines are redshifted to the near-infrared (NIR) which has a high and variable sky background. In contrast to actively star-forming galaxies, measuring a stellar-velocity dispersion of quiescent galaxies requires a continuum detection, with moderate signal-to-noise ratios (S/N). Because of the long integration times required to achieve the necessary S/N enabling a stellar velocity dispersion measurement, the community has been restricted to observing the brightest galaxies at these redshifts \citep{onodera2010, vandesande2011, toft2012, vandesande2013, bezanson2013b, belli2014a}, which \citet{vandesande2014} showed is also biased towards younger post-starburst galaxies.

In order to probe higher redshifts and/or lower masses, and circumvent the need for long integration times prior to the era of the James Webb Space Telescope (JWST), we aim to take advantage of the brightening and magnifying effects of strong gravitational lensing. This tool has been successfully implemented in studying the properties of distant star-forming galaxies, with higher resolution and better signal-to-noise than normally possible including lyman-break \citep[e.g.][]{smail2007}, sub-millimetre  \citep[e.g.][]{vieira2013} and UV-bright galaxies \citep[e.g.][]{brammer2012,sharon2012,vanderwel2013}. We aim to extend the utility of gravitational lensing to red, quiescent galaxies. 

Lensed, quiescent galaxies at high redshifts ($z>2$) are comparably more difficult to find than lensed star-forming galaxies for a variety of reasons. First quiescent galaxies show a declining number density with increasing redshift \citep[i.e.,][]{muzzin2013b}. Thus, there are fewer quiescent galaxies to be lensed at high redshift as compared to star-forming galaxies. Secondly, blue, lensed star-forming galaxies stand out in red, quiescently dominated galaxy clusters, whereas red, lensed galaxies do not. One of the best places to search for lensed galaxies is behind galaxy clusters as they have deep potential wells. As a result of high star formation rates (SFR), SMGs exhaust their gas on relatively short timescales, and are thus extremely rare in local galaxy clusters. Because of SMG rarity in local clusters, the foreground lensing cluster has few sources in the sub-mm images allowing for trivial detection of the lensed SMGs. This is not the case for quiescent galaxies, where the foreground cluster is also NIR bright. Blue, star-forming galaxies are UV-bright, and lensed candidates at redshift $\sim2$ have this emission shifted to the rest-frame optical, making the high-redshift blue galaxies behind clusters optically bright. With the existence of wide and relatively deep optical surveys such as SDSS, there is a wealth of lensed blue galaxies \citep[e.g,][]{stark2013}, however large area NIR surveys of comparable depth are not available. 

As such, there are only five red, lensed galaxies presented in the literature. \citet{auger2011} present an intermediate redshift ($z=0.6$) lensed candidate which is multiply imaged. Two of the high-redshift ($z=1.71,2.15$) examples in the literature \citep{geier2013} are singly imaged, which are more difficult to create lens models for.  There are only two examples of multiply imaged red-lenses at high redshift. One, found by \citet{newman2015}, with a spectroscopic redshift of $z=2.636$, and the other is the object of this study which was  first identified by \citet{muzzin2012}. 

In this paper we present X-Shooter spectroscopy, and a stellar velocity dispersion measurement of $\mathrm{COSMOS~0050+4901}$, a quiescent galaxy found by \citet{muzzin2012} in the COSMOS/UltraVISTA field \citep{mccracken2012}. With the current data, this is now the highest redshift quiescent galaxy with a stellar velocity dispersion measurement, as well as the least massive quiescent galaxy beyond redshift 2 with a rest-frame optical spectrum. 

We assume a $\mathrm{\Lambda}$-CDM cosmology ($H_{0}=\mathrm{70~kms^{-1}Mpc^{-1}}$, $\Omega_{M}=0.3$, and $\Omega_{\Lambda}=0.7$), and AB magnitudes. 

\section{Data}
\label{sec:data}

%Image of galaxy
\begin{figure}
  \begin{center}
  \includegraphics[width=\linewidth]{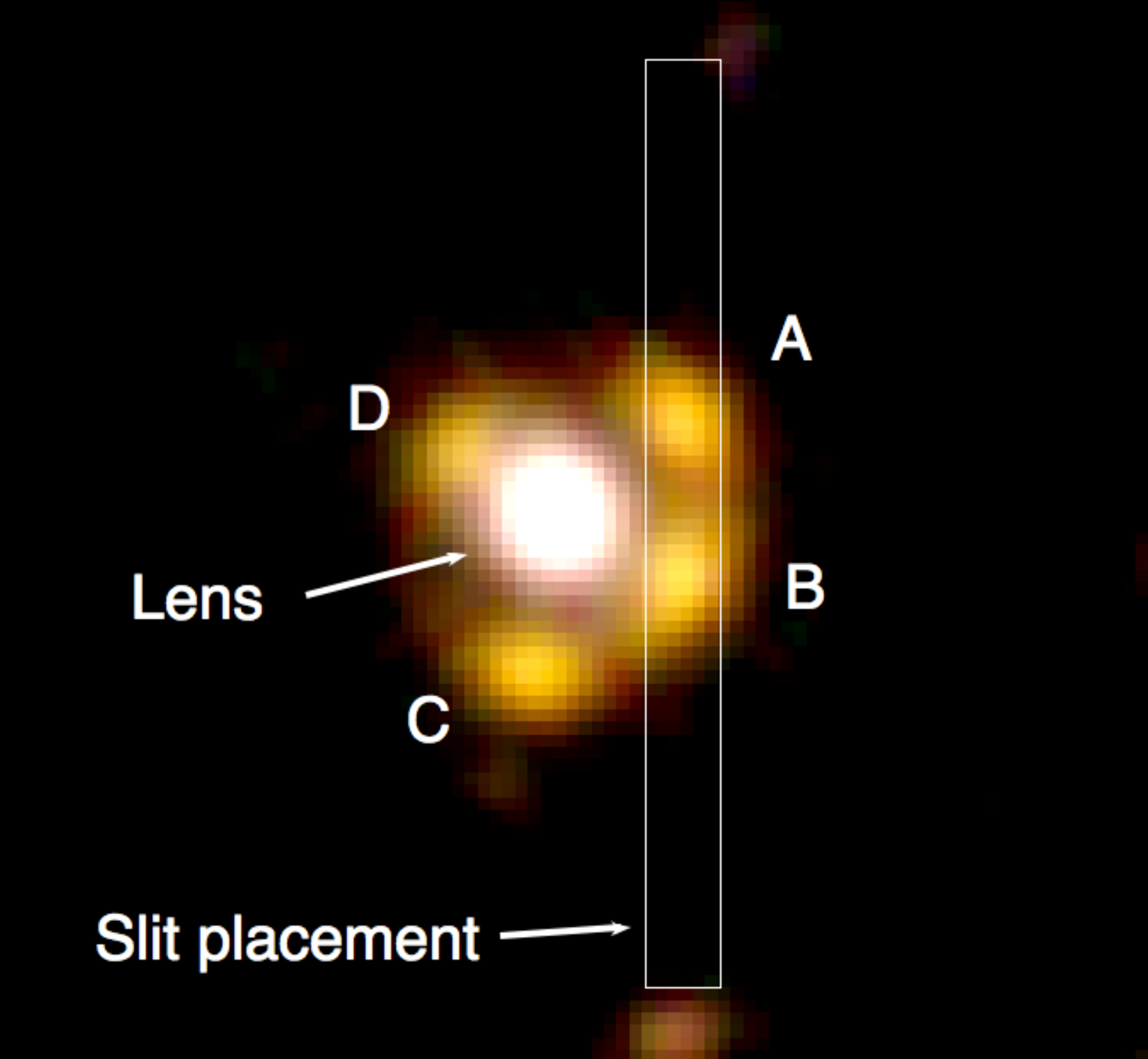}
  \caption{RGB-image using UltraVISTA Ks, H, and J-band DR2 images. The source images are labeled A, B, C, and D, and the foreground lens is indicated. The source images are noticeably redder than the lens because of the discordant redshifts ($\mathrm{z_{lens}}=0.960$ and $\mathrm{z_{source}}=2.76$) which result in the continuum emission from the lens and source galaxies peaking in different bands. The position of the X-Shooter slit on the sky is indicated. Note that the slit was placed such that it falls on galaxies A and B.}
  \label{fig:hband}
  \end{center}
\end{figure}

\subsection{COSMOS 0050+4901}
$\mathrm{COSMOS~0050+4901}$ is a strongly-lensed system where the lens is a single galaxy at $z=0.960$ found serendipitously in the COSMOS/UltraVISTA field \citep{mccracken2012} as a group of exceptionally bright, red galaxies. The source is quadruply imaged, with photometric redshifts of $z\sim2.3-2.4$ (depending on which of the multiple images is analyzed; \citealt{muzzin2012}, hereafter M12). The brightest 3 images are magnified by a factor of $\sim5$ (M12, Muzzin et al. in prep). In Fig.~\ref{fig:hband} we show a $3$-color UltraVISTA J, H and  Ks-band image of the lens-source system. Included in this figure is the placement of the slit used for the spectroscopic data (see Sec.~\ref{sec:spectra}). As illustrated by the slit position shown in Fig.~\ref{fig:hband}, we have obtained spectroscopy of two images, effectively doubling our exposure time. Fig.~\ref{fig:hband} also qualitatively illustrates the difference in color between the foreground lens and the source images as a result of their differing redshifts.  

M12 performed an initial estimate of the structural properties of the galaxy using the ground-based $\mathrm{K}_{s}$-band data. Assuming the best photometric redshift at that time, they estimated an $r_{e}=0.64^{+0.08}_{-0.18}~\mathrm{kpc}$ and a Sersic index of $n=2.2^{+2.3}_{-0.9}$. Recently, we have obtained high-resolution HST~F160W imaging of COSMOS~$0050+4901$. With the deep HST image and spectroscopic redshift determined in this analysis, Muzzin et al. (in prep) determined a well-constrained lens model, which allowed them to accurately determine the circularized $r_{e}$ (corrected for magnification) and $n$ to be $0.86^{+0.19}_{-0.14}~\mathrm{kpc}$ and $3.50^{+0.68}_{-0.60}$.

\subsection{Rest-Frame $UVJ$ Colors}
\label{sec:color}

\citet{williams2009} demonstrate that galaxies display a clear bi-modality in $U-V$ and $V-J$ color-space out to $z=2$. Galaxies tend to separate into two sequences in color-color space, one consisting primarily of star forming galaxies, and one primarily consisting of quiescent galaxies. This bi-modality is driven by a galaxy's UV+IR determined SFR. Beyond $z=2$, measurement errors, and completeness limits reduces this bi-modailty \citep{williams2009, muzzin2013b}. 

We calculate the rest-frame $U-V$ and $V-J$ colors for $COSMOS~0050+4901$ using the photometric data presented in M12 (via \textit{EAZY}; \citealt{brammer2008}). We de-blending the lens from the images, which showed non-negligible contamination in the PSF-matched ground based imaging, by simultaneously fitting both the images and the lens using \textit{GALFIT} \citep{peng2010} (further details may be found in M12).

\begin{figure}
  \begin{center}
  \includegraphics[width=\linewidth]{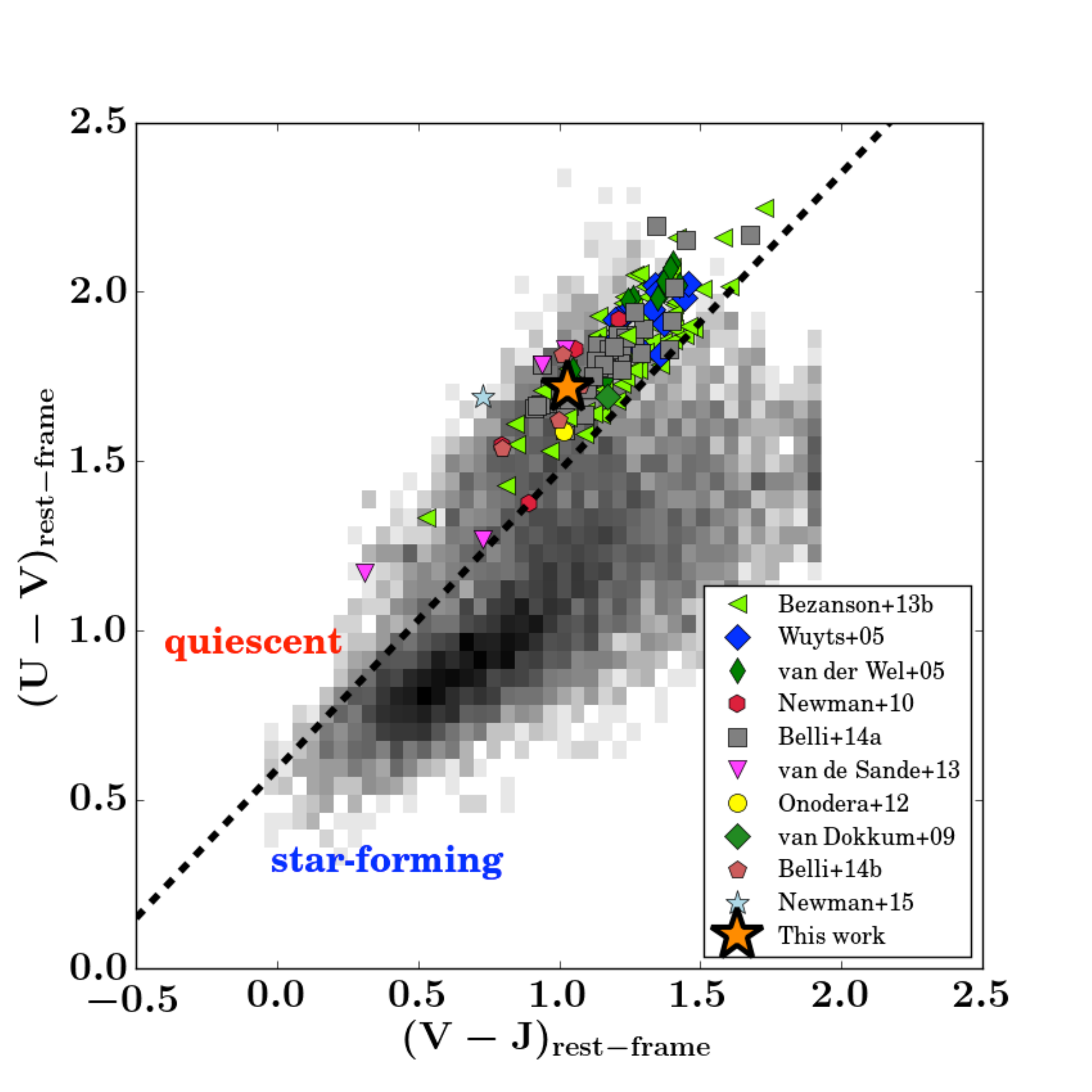}
  \caption{Rest-frame \textit{U-V} color vs \textit{V-J} color. Grey points are a redshift selected ($1.5<z<2.5$) comparison sample from the $\mathrm{K_{s}}$-band selected UVISTA catalog \citep{mccracken2012}. The black dashed line is the colour selection implemented by \citet{vandesande2015} to separate quiescent and star forming galaxies. Coloured points are a high redshift ($0.6<z<2.3$) sample compiled by \citet{vandesande2015}, with the literature sources indicated in the plot legend. Our object (orange star) sits on the UVISTA $1.5<z<2.5$ red sequence, and within the locus of other high redshift, quiescent galaxies with measured stellar velocity dispersions.}
  \label{fig:uvj}
  \end{center}
\end{figure}

In Fig.~\ref{fig:uvj} we show that our object falls on the quiescent region in the $UVJ$ diagram (using $z_{spec}=2.756$; the determination of which is described in Sec.~\ref{sec:redshift}). We compare and contrast it to a quiescent, spectroscopic sample compilation (from \citealt{vandesande2015}, with a redshift distribution of $0.6<z<2.2$). This sample contains 63 galaxies at $0.4<z<1.6$ from \citet{bezanson2013b}, 38 galaxies at $1<z<1.4$ from \citet{belli2014a}, 18 galaxies at $0.6<z<1.1$ from \citet{vanderwel2005}, 16 galaxies at $z\sim0.8$ from \citet{wuyts2004}, 5 galaxies at $1.2<z<1.6$ from \citet{newman2010}, 4 galaxies at $1.4<z<2.1$ from \citet{vandesande2013}, 3 galaxies at $2.1<z<2.4$ from \citet{belli2014b}, 1 galaxy at $z=2.2$ from \citet{vandokkum2009}, 1 galaxy at $z=1.8$ from \citet{onodera2012}, and 1 galaxy at $z=2.6$ from \citet{newman2015} (see Table in \citealt{vandesande2015} for further details). As stated in \citet{vandesande2015}, the sample is selected based on the availability of kinematic measurements in the literature. Thus, this sample is biased towards brighter objects. In Fig.~\ref{fig:uvj} we also indicate the UVJ color selection from \citet{vandesande2015} as the dashed-black line. 

Also plotted in grayscale in Fig.~\ref{fig:uvj} is a redshift-selected ($1.5<z<2.5$), photometric sub-sample from the K-band selected catalog of UltraVISTA from \citet{muzzin2013a} (with the limiting magnitude $K<24.4$ in a $2.1^{\prime\prime}$ aperture). This sub-sample contains both star forming, and quiescent galaxies. The redshift range was chosen in order to highlight the color bi-modality, which as previously mentioned, is erased at higher redshifts due to incompleteness and measurement errors. In comparison to both the spectroscopic, and photometric samples, our object has colors similar to galaxies with quiescent populations.

%UVJ diagram of high redshift galaxies

\subsection{Spectroscopic Data}
\label{sec:spectra}
Data were obtained using the X-Shooter instrument on the VLT UT2 \citep{dodorico2006,vernet2011} with the K-band blocking filter in place. The target was observed in service mode; the observations were carried out between 2012 December and 2014 February (program Muzzin 090.B-0452(A), and DDT program Muzzin 288.B-5043(A)). All observations had clear sky conditions and an average seeing of $0.8^{\prime\prime}$. A $0.9^{\prime\prime}$ slit was used in the NIR, aligned in the North-South direction with two of the images on the slit, as shown in the UltraVISTA color image in Fig.~\ref{fig:hband}. X-Shooter simultaneously takes spectra in the observed UVB and VIS. The UVB and VIS arm data had no signal, as expected from the very red SED.

The NIR sky changes on short timescales ($\sim\mathrm{minutes}$) and to compensate for changing sky levels, it is customary to perform a nodding pattern, with two frames adjacent in time subtracted from each other with the object offset in adjacent frames, often referred to as an ABBA observing pattern. Given the size of the X-Shooter slit ($0.9^{\prime\prime}$x$11^{\prime\prime}$), the spatial extent of our object was such that there was insufficient space to perform this nodding pattern with enough empty sky for a successful sky subtraction. As such, observing blocks were 10 minutes, with a nodding pattern offset in declination in $0.4^{\prime\prime}$ increments to a maximum offset of $2^{\prime\prime}$ which corresponds to a shift on the slit of between 2 and 10 pixels. These offsets were made for the identification and removal of bad pixels. 

The images were reduced using the method most commonly used in optical spectroscopy (see Section~\ref{sec:reduction}). Of 64 science frames with 600s exposure of exposure time on each frame, 5 frames did not contain the target and were thus not used in the final combination (but utilized in sky subtraction - see Section~\ref{sec:reduction}). This results in a science image with a total exposure time of 9.8hrs. However, obtaining a similar S/N spectrum on a single galaxy with comparable un-lensed magnitude ($\mathrm{K}\sim22.7$) would require 215 hours when accounting for the fact that our observations are sky-limited, 2 objects fall on the slit and that we are only semi-resolved. This gain in S/N demonstrates the substantial observing advantage provided by strong gravitational lensing. Additionally, a B9V telluric standard star (Hipparcos~049704) was also observed before and after the science target for removal of atmospheric absorption lines, as well as relative flux calibration between orders.  

\subsection{Spectroscopic Reduction}
\label{sec:reduction}
The data were reduced with the ESO pipeline for X-Shooter (ver 3.10; \citealt{modigliani2010}), following the ``physical" mode reduction chain using EsoRex. Individual frames were reduced in stare mode, as the extent of the object on the slit made standard sky subtraction difficult. Bad-pixel masks were generated using IRAF task \textit{ccd\_mask}, and bad-pixels corrected for using the IRAF task \textit{fixpix}.

Because of the methodology of our sky subtraction, we found several detector artefacts on the images which complicated this procedure. In order to subtract these artefacts, we generated a sky frame out of 5 blank sky frames from our observations, as well as other 10 min exposures which used the K-band blocking filter found in the X-Shooter archive. We used a total of 28 frames. These frames were all median combined to generate a high S/N sky-frame. This sky-frame was then subtracted from the science frames to subtract the detector artefacts.

The OH-emission lines in the NIR vary in flux, and change on short time scales, so the sky-frame subtraction could not account for sky lines. To account for this, the sky was modelled along each column in the spatial direction, while masking out rows which contained galaxy flux. This modelled sky was then subtracted from each column. The individual exposures were then median combined order by order. 

The telluric standard spectra were reduced in the same way as the science frames. We constructed a response spectrum from the telluric star, and a black body curve with a $T_{eff}$ matching a B9V star. Residuals from Balmer absorption in the telluric standard were removed by interpolation. The science observations were corrected for instrumental response and atmospheric absorption by division of the response spectrum.

To extract the spectrum, a 1D light profile was fit to each wavelength pixel (or column) along the spatial direction. The light profile was modelled from a median combination of all these fitted profiles from an order in the H-band (the highest S/N region of our spectrum). Order number 17 is shown in Fig~\ref{fig:redux} to illustrate. The light profile found in the top panel of Fig.~\ref{fig:redux} was fit, with a background term, to each column of the combined, 2D spectrum:

%reduction methodology example
\begin{figure*}
  \begin{center}
  \includegraphics[width=\textwidth]{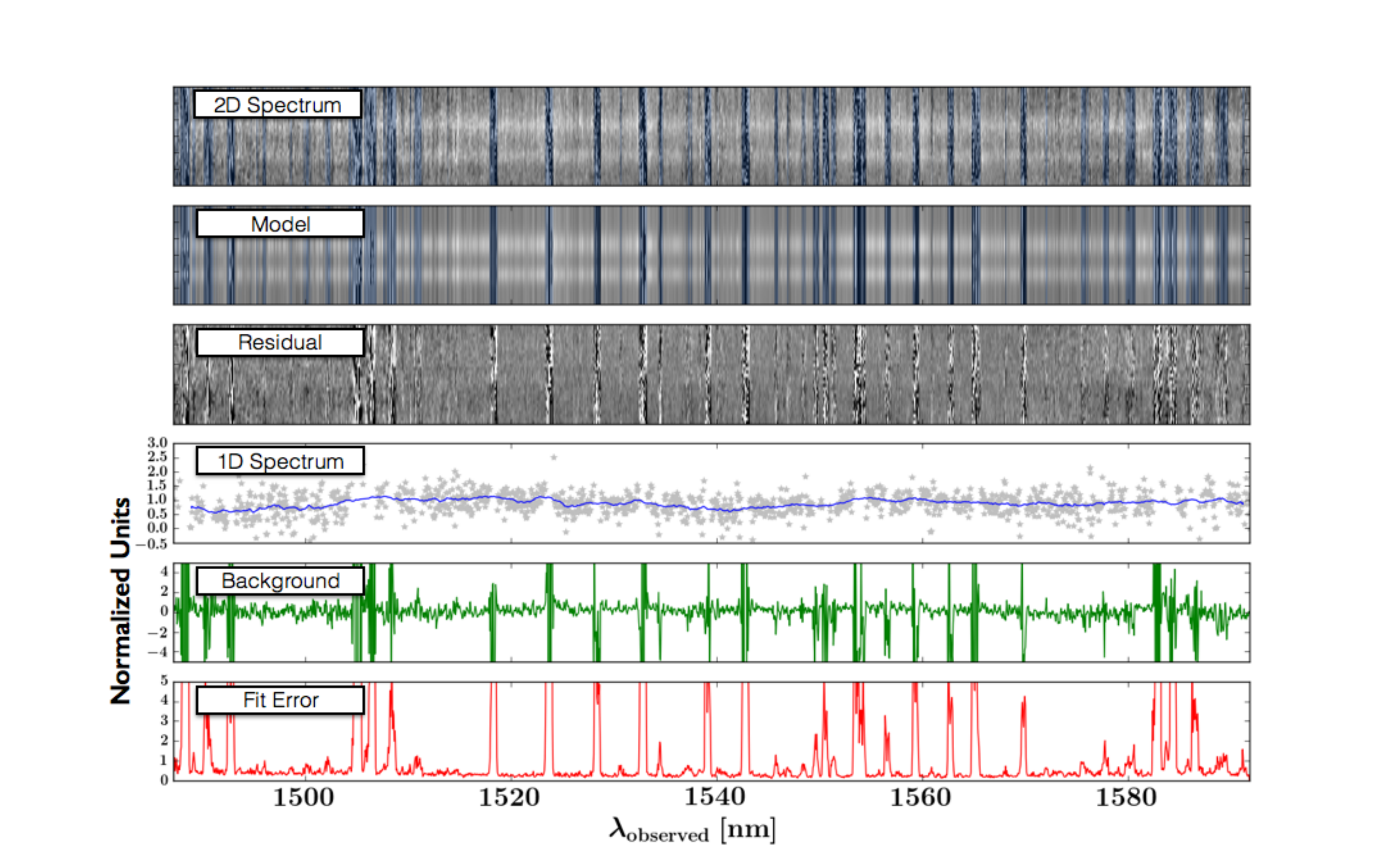}
\caption{Above is an example of our prescription for column rejection and spectral extraction. From to to bottom: \textbf{(1)} The median combined 2D spectrum \textbf{(2)} The model reconstruction from the double gaussian fitting as described in ``Spectroscopic Reduction'' \textbf{(3)} The residuals of (1) and (2) \textbf{(4)} The coefficient of the relative strength of the double-peak profile (effectively the extracted 1D spectrum with normalized units) \textbf{(5)} The coefficient of the background term \textbf{(6)} The relative error error from the fit. The columns are rejected above a specified `Fit Error' which varied with each spectral order. The blue shaded regions in (1) and (2) are the columns which are rejected.}
\label{fig:redux}
\end{center}
\end{figure*}

\begin{equation}
c_{\lambda} = a_{\lambda}P_{y} + b_{\lambda}
\label{eq:red_line}
\end{equation}

$\lambda$ and $y$ refer to the spatial and spectral dimensions. $c_{\lambda}$ refers to the column, $P_{y}$ is the double peaked profile fit to each column, $a_{\lambda}$ and $b_{\lambda}$ are the fitted coefficients. $a_{\lambda}$ is effectively the 1D spectrum, and $b_{\lambda}$ is the background term (see 4th and 5th panels in Fig.\ref{fig:redux}). The error spectrum (bottom panel of Fig.~\ref{fig:redux}) is the covariance of the fit, using the $1\sigma$ value of the pixels at each location. Skylines and bad columns were flagged using an error cutoff (above which the columns would be rejected). 

\begin{figure*}
  \begin{center}
  \includegraphics[width=\textwidth]{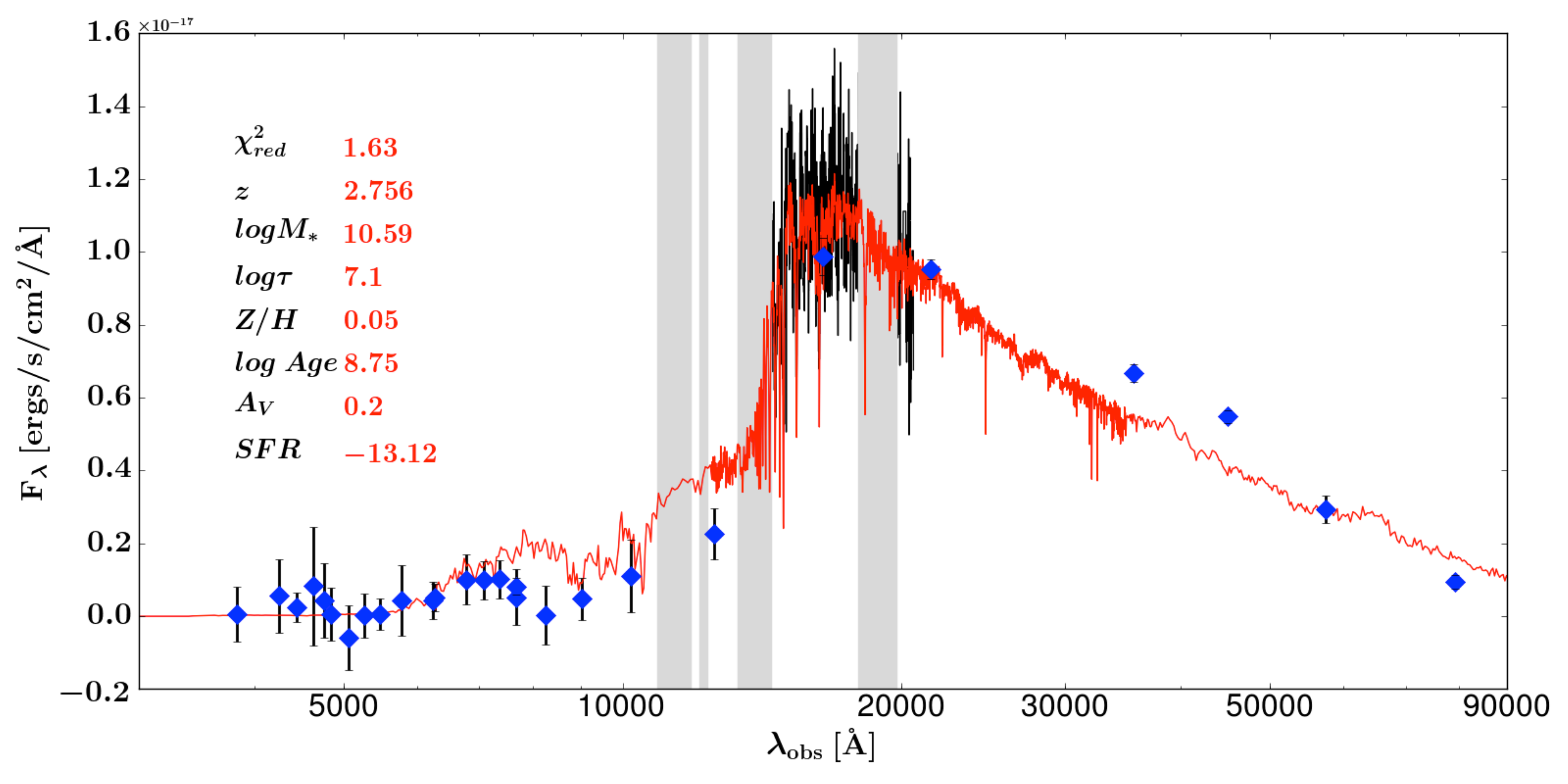}
\caption{Best-fit FAST BC03 template (red-line) with photometry (blue points) and H-band X-Shooter spectrum (black-line). Grey regions indicate strong atmospheric absorption resulting in low S/N spectra. The J-band spectrum was omitted due to contamination from the lensing galaxy (see text for further discussion). Best-fit stellar population parameters and their 68\% confidence intervals can be found in Table~\ref{tab:fast}.}
\label{fig:zoomout}
\end{center}
\end{figure*}

\begin{figure*}
  \begin{center}
  \includegraphics[width=\textwidth]{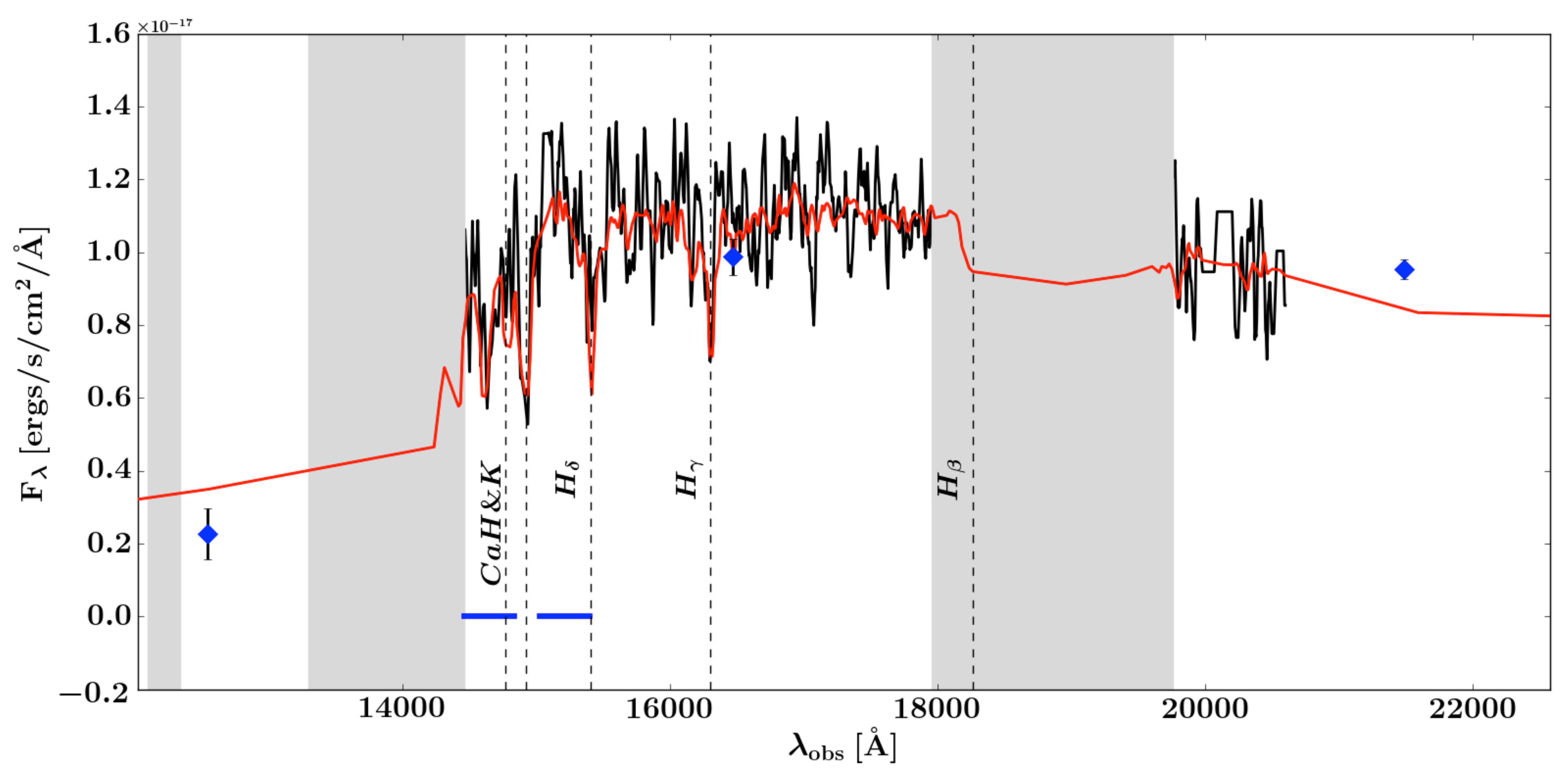}
\caption{Similar to Fig.~\ref{fig:zoomout}, but with the wavelength range restricted to the H-band, and the location of prominent absorption lines indicated. The fitted spectrum was smoothed using a bi-weight mean with a bin size of 10 good points (i.e. those that were not rejected by our criteria described in Sec.~2.3). The spectrum shown above has been smoothed to better illustrate the presence of absorption lines.}
\label{fig:zoomin}
\end{center}
\end{figure*}

The low-resolution spectrum was constructed by binning the 1D spectrum in the wavelength direction, using a bi-weight mean (as described in \citealt{vandesande2013}) with a bin size of 10 good pixels. We show the spectrum, along with photometry and best fit BC03 model in Fig.~\ref{fig:zoomout}, and~\ref{fig:zoomin}. 

%full spectrum with best fit fast and photometry

%full spectrum with best fit fast and photometry (zoomed in)

\section{Structural Properties and Stellar Populations}
\label{sec:prop_and_pop}

\subsection{Redshift Determination}
\label{sec:redshift}
A spectroscopic redshift was measured using FAST \citep{kriek2009}, an IDL-based fitting routine which fits stellar population synthesis models to photometry/spectra. This fitting is described in more detail in Section~\ref{sec:stellpop}.

Our best-fit redshift is $z_{\mathrm{spec}}=2.756\pm0.001$. This is a high-confidence measurement, as numerous absorption lines, such as $\mathrm{Ca~H\&K}$, $\mathrm{H}\delta$, and $\mathrm{H}\gamma$ are detected. Notably, the difference between the photometric ($\mathrm{z_{phot}}=2.4\pm0.13$) and spectroscopic redshift ($z_{\mathrm{spec}}=2.756\pm0.001$) is more than twice the uncertainty on the photometric redshift. The possible explanations for this discrepancy are discussed below. 

The difference between the $z_{\mathrm{spec}}$ and $z_{\mathrm{phot}}$ might originate from mistaking the $4000\mathrm{\AA}$-break and Balmer-break. The $4000\mathrm{\AA}$-break and Balmer-break ($\sim3640\mathrm{\AA}$) are two continuum features which are difficult to resolve in photometric datasets due to their proximity in wavelength. Both continuum features result in a color differential between each side of the $4000\mathrm{\AA}$/$3650\AA$ spectral region, the strength of which correlates with age in both cases. They do, however, originate from different physical processes, and thus \textit{how} they correlate with age is different. The $4000\mathrm{\AA}$ break is a result of absorption by ionized metals which is strongest in older and high-metallicity stellar populations. The Balmer-break marks the limit of the Balmer series and blending of higher order Balmer lines, and is strongest in A-stars. The strength of the Balmer-break monotonically increases before peaking at intermediate ages ($\sim0.5-1$~Gyr) when the stellar light is dominated by A-stars. 

Both these features can be prominent in the continuum of quiescent galaxies, and fall into the NIR between $1.5<z<3$. The discrepancy between the $z_{phot}$ and $z_{spec}$ determination is likely caused by a combination of wavelength gaps between the transmission curves in the NIR bands (J,H, and $\mathrm{K_{s}}$), as well as wide bandwidth in the same filters. This results in a J-H color appropriately explained by either the presence of the $4000\mathrm{\AA}$-break at lower z, or the Balmer-break at higher z. 

This uncertainty associated with the breaks falling between the NIR filters likely caused the fitting confusion between a quiescent galaxy with a prominent $4000~\mathrm{\AA}$ break at $z=2.4$, and a post-SB galaxy with a $3650~\mathrm{\AA}$ Balmer-break at $z=2.8$. This illustrates the challenges of using $\mathrm{z_{phot}}$ only. This issue is well-known, and that even with high S/N NIR photometry, photometric redshifts at $z>2$ can be uncertain due to these issues. This has led to the use of NIR ``medium bands" as in the NEWFIRM Medium Band Survey \citep{whitaker2011} and zFOURGE Survey \citep{straatman2014} to provide improved photometric redshifts for galaxies where the Balmer and $4000\AA$-breaks fall in the observed NIR.

This redshift difference is large enough that it changes the stellar population parameters by M12 at a level that is larger than the quoted uncertainties in that paper, particularly the stellar mass, age, and dust content. In the next section we present revised values for these parameters using the spectroscopic redshift. 

\subsection{Stellar Population Properties}
\label{sec:stellpop}
Stellar population properties are estimated using FAST \citep{kriek2009}. We used \citet{bc2003} templates, with a delayed, exponentially declining star formation history (with timescale $\tau$), a \citet{chabrier2003} IMF, and \citet{calzetti2000} dust law.

We have simultaneously fit both the photometry and H-band spectrum. We have omitted the J-band spectrum but include the J-band photometry. The spectrum in the J-band is very low S/N as the flux from the lens peaks in the J-band (see Fig.~4 of M12). The image and lens are spatially close with contamination affecting the continuum strength. The best-fit parameters are summarized in Table~\ref{tab:fast}. The best-fit stellar population parameters provide a best-fit FAST age $\log{\mathrm{Age/yr}}=8.75^{+0.07}_{-0.07}$, stellar-mass $\log{\mathrm{M_{*}/M_{\odot}}}=10.59^{+0.04}_{-0.05}$ (corrected for lensing), and $A_{V}=0.2^{+0.26}_{-0.20}$. This galaxy appears post-starburst, and is striking in that even at $z=2.8$, intermediate-mass galaxies with quiescent stellar populations exist. 

\citet{muzzin2013b} suggests that the quenched fraction for galaxies of $\mathrm{\log{M_{*}/M_{\odot}}}>10.8$ is $\sim20\%$ at these redshifts. The identification of galaxies with quiescent populations in \citet{muzzin2013b} is based on rest-frame color selection and  $\mathrm{z_{phot}}$. Here we confirm unambiguously through spectroscopy that these galaxies do exist at these redshifts. 

Compared to the values of M12, the effects of fitting the spectrum and photometry with the spectroscopic redshift result in a best-fit where the age changes from $\log{\mathrm{Age/yr}}=9.0^{+0.2}_{-0.2}$ to $\log{\mathrm{Age/yr}}=8.75^{+0.07}_{-0.07}$, stellar-mass from $\log{\mathrm{M_{*}/M_{\odot}}}=10.82^{+0.05}_{-0.07}$ to $\log{\mathrm{M_{*}/M_{\odot}}}=10.59^{+0.04}_{-0.05}$, and dust from  $A_{V}=0.9^{+0.2}_{-0.6}$ to $A_{V}=0.2^{+0.26}_{-0.20}$. It is important to note that these values do not agree (within the $1\sigma$ errors) with the values reported by M12. However M12 underestimated the uncertainties associated with the $z_{phot}$, leaving a deficit in the error budget resulting in a disagreement of values. With the addition of a $z_{spec}$, the best-fit galaxy is younger, less massive, and contains less dust than previously determined by M12.

In order to understand the differences in best-fit stellar population parameters between M12 and the present study, we re-fit the data using the photometry and the spectroscopic redshift (omitting the spectra). This produced a different set of parameters from our best-fit and closer to the age, mass and dust content of M12, suggesting that the spectrum does drive the fit. We conclude that both the spectroscopic redshift and the spectrum itself, which shows strong Balmer absorption, drive the changes in the stellar populations. 

We note that in Fig.~\ref{fig:zoomout}, it is clear that our best-fit to the photometry and spectroscopy is not ideal. The most striking mismatch occurs in the IRAC bands. The disagreement between the spectrum and photometry in the far-infrared could be attributed to the challenges associated with de-blending the source from the lens, and the lens images from one another. In the IRAC bands, the FWHM of the PSF becomes comparable to the separation between source galaxies and the lens which is $2^{\prime\prime}$ for this system. Accurately separating the flux becomes more difficult than in the observed optical and NIR where the PSF is smaller. We ascertained the effect of the IRAC bands on the fit by re-fitting the spectrum and photometry without IRAC. We found the stellar population parameters to be the same within $1\sigma$ and conclude that the IRAC bands do not strongly influence the outcome of the fitting. The current analysis now includes age sensitive absorption features (see below), as well as the new spectroscopic redshift, and therefore we are confident in the stellar population parameters and associated uncertainties determined in this study. 

In Fig.~\ref{fig:zoomin}, we find a weak or absent Ca K absorption in the data, whereas the model suggests a stronger absorption line. The observed wavelength of Ca K at z=2.756 is 4780~$\AA$ which overlaps with a strong sky-line at 14793~$\AA$ leading to poor spectral extraction in that region. Thus, the mismatch between the data and model Ca K absorption strength is likely caused by poor data quality in that region. We note that regions affected by strong skylines have larger errors, and will therefore have lower weight in the full spectral fitting.

\begin{deluxetable*}{lcccccccc}
\tabletypesize{\scriptsize}
\tablecaption{Stellar Population Synthesis Properties}
\tablewidth{0pt}
\tablehead{
\colhead{$\mathrm{Z}$} & \colhead{$\mathrm{z_{spec}}$} & \colhead{log~$\tau$} & 
\colhead{log~Age} & \colhead{Av} & \colhead{log~$M_{*}$} &\colhead{log~$\mathrm{SFR^{\dagger}}$} &\colhead{log~sSFR} & \colhead{$\chi_{\mathrm{red}}^{2}$}
\\
\colhead{} & \colhead{} & \colhead{(yr)} & 
\colhead{(yr)} & \colhead{(mag)} & \colhead{($M_{\odot}$)} &\colhead{($M_{\odot}~\textrm{yr}^{-1}$)} &\colhead{($\mathrm{yr}^{-1}$)} &\colhead{}
}
\startdata
$0.050$ & $2.756^{[2.757]}_{[2.755]}$ & $7.1^{[7.63]}_{[7.00]}$ & $8.75^{[8.82]}_{[8.68]}$ & $0.2^{[0.46]}_{[0.0]}$  & $10.59^{[10.63]}_{[10.54]}$  & $-13.12^{[-2.4]}_{[-17.
52]}$ & $-23.71^{[-13.00]}_{[-28.07]}$  & $1.63$ 
\\
&  &  &  &  & ($11.25$\tablenotemark{*})  & ($-12.46$\tablenotemark{*})  &   &  

\enddata
\tablenotetext{$\dagger$}{This is from the 30-band SED fit with a $\tau$-model star formation history, and is effectively a UV-dust-corrected SFR.}
\tablenotetext{*}{FAST output before adjusting for lensing magnification}
\tablecomments{The best-fit FAST parameters and their values within 68\% confidence intervals, adjusted for the lensing magnification (where appropriate) from \citealt{muzzin2012}}
\label{tab:fast}
\end{deluxetable*}

In addition to the stellar population parameters fit with FAST, we measured the Lick index $\mathrm{H}\delta_{A}$ \citep{worthey1997} and $\mathrm{D}_{n}(4000)$ \citep{balogh1999} break, which are features shown to be sensitive to age \citep{kauffmann2003}. With an $\mathrm{H}\delta_{A}$ measurement from our spectrum, as well as coverage of the $4000\mathrm{\AA}$ break (as seen by the blue horizontal bars in Fig.~\ref{fig:zoomin}), we are able to independently verify our model age determination. This independent age verification from the absorption features is important because of inherent degeneracies in fitting the SED. In Fig.~\ref{fig:dn4000_hdelt}, we have plotted $\mathrm{H}\delta_{A}$ as a function of $\mathrm{D}_{n}(4000)$ of our object, as well as a random sample of SDSS galaxies which contain both star-forming and quiescent galaxies. 

Over-plotted in Fig.~\ref{fig:dn4000_hdelt} are three different model tracks generated using GALAXEV \citep{bc2003}. The best-fit super-solar metallicity of $Z=0.05$ from FAST was used in each model track. Two fiducial models (a burst, and constant star formation history) are plotted in blue and magenta to highlight the extremes in the parameter space. A model with a delayed exponential SFH is also plotted in red, using the best-fit parameters from Table~\ref{tab:fast}. The red point corresponds to the age of our best fit using FAST. The separation between the $\mathrm{H}\delta_{A}$ vs. $\mathrm{D}_{n}(4000)$ determined age and FAST modelled best-fit age is small, confirming the FAST best-fit age in a model independent way.  Fig.~\ref{fig:dn4000_hdelt} also shows very little difference in the $\tau$ vs. delayed-$\tau$ models, and that the star formation history is dominated by a population which is consistent with a single burst. From Fig.~\ref{fig:dn4000_hdelt}, we confirm that this galaxy is indeed relatively young, post-starburst, and consistent with being recently quenched.

\subsection{MIPS 24~$\micron$ Photometry}
\label{sec:twentyfourmicron}

As described in Sec. 5.3 of M12, there is corresponding MIPS $24~\micron$ data which was remarked upon. We briefly summarize their findings. They detected observed-frame $24~\micron$ emission at  $4\sigma$ in the vicinity of the lens system. However, they found the MIPS source to be offset to the south-west by several arcseconds which suggests that the lens system is not the correct counterpart. Under the possibility that the MIPS detection \textit{is} coincident with the lens-source system, M12 determined an estimate of the sSFR. Since the FWHM of the PSF is $5.5^{\prime\prime}$, individual sources could not be resolved in MIPS. Photometry was therefore performed in a $7^{\prime\prime}$ aperture which surrounded the entire lens system. For the source and lens, at $z_{phot}=2.4$, they find $\mathrm{\log{sSFR}}=-9.93^{+0.20}_{-0.20}$ which is below the star-forming main sequence for their derived stellar mass at $z_{phot}=2.4$. Thus, even in the scenario where all of the MIPS emission is associated with the lensed galaxies, they would still be classified as quiescent. 

\begin{figure}
  \begin{center}
  \includegraphics[width=\linewidth]{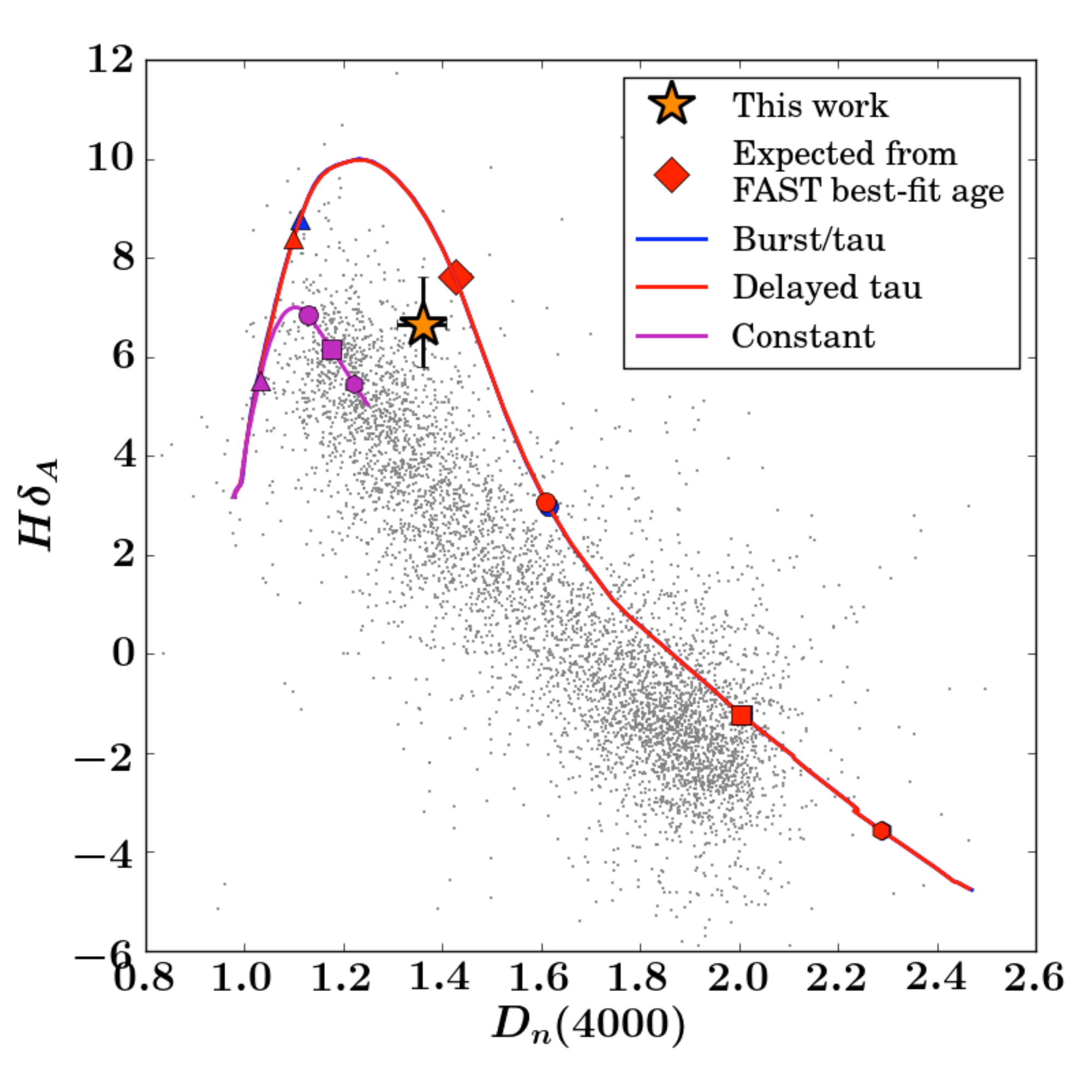}
\caption{$\mathrm{H}\delta_{A}$ as a function of $\mathrm{D}_{n}(4000)$. Gray points are a random sample of SDSS galaxies. The orange star is the measurement from the H band spectroscopy. The solid, colored lines are different \citet{bc2003} models with different SFH (burst, delayed exponential and constant) with the best fit values from Table~\ref{tab:fast} used as inputs. Different benchmark ages are indicated along each track with a triangle ($0.1~\mathrm{Gyr}$), circle ($1.0~\mathrm{Gyr}$), square ($3.0~\mathrm{Gyr}$), and hexagon ($10.0~\mathrm{Gyr}$). Marked on the delayed exponential SFH track (the model of choice from FAST), the best-fit age is indicated (red diamond). The close separation reaffirms our age determination.} 
\label{fig:dn4000_hdelt}
\end{center}
\end{figure}

The new $z_{spec}$ determination will effect the sSFR found by M12, which we re-calculate below. We follow the same procedure to find a total un-lensed mass of the source galaxies of $\log{M_{*}/\mathrm{M_{\odot}}}=11.56^{+0.08}_{-0.04}$. With the photometry from M12, and the templates of \citet{dale2002}, the implied un-lensed SFR is $370^{+96}_{-83}~M_{\odot}/yr$. This yields a $\log{sSFR}=-8.99^{+0.12}_{-0.12}$. The sSFR of a star-forming main sequence galaxy is $\mathrm{\log{sSFR}}\sim-8.6$ for $\log{M_{*}/\mathrm{M_{\odot}}}=10.59$ between $2.5<z<3.5$ \citep{schreiber2015}. This implies that if the MIPS detection is associate with the lensed system, then this galaxy is only 0.3 dex below the star-forming main sequence. We find this implied level of star formation unlikely for two reasons. The first is that the MIPS and NIR sources are offset from one another, and the MIPS detection is not likely associated. The second is our model independent age determination via the strengths of $\mathrm{H}\delta_{A}$ and $\mathrm{D}_{n}(4000)$ (as seen in Fig.~\ref{fig:dn4000_hdelt}) which emphasize an older age for the majority of the stars in this galaxy.

If the MIPS detection were coincident then all star-formation would need to be dust-enshrouded, and very recent, so that no young stars are visible outside the birth-clouds. An alternative explanation is that the lens-source system could host an AGN, but we do not see emission lines in the near-IR or optical. Additionally, M12 looked for X-ray and radio detections in \textit{XMM-Newton} and the Very Large Array observations of the COSMOS field and found no detection in the vicinity of the system. 

%hdelta vs dn4000

\subsection{Stellar Velocity Dispersion}
\label{sec:dispersion}

The presence of strong absorption features provide us with the means to study the kinematics of this galaxy, and determine a stellar velocity dispersion. With this measurement, we are able to calculate the dynamical mass and place strong constraints on the baryonic contribution to the total mass budget. Although random errors on broadband photometry can be quite low with state-of-the-art instruments, systematic uncertainties in mass determinations are difficult to estimate accurately \citep{conroy2009}. Thus, velocity dispersions are key for placing upper limits on the total mass of the system, and hence validating stellar mass estimates. 

A stellar velocity dispersion was estimated using Penalized Pixel-Fitting (pPXF) \citep{cappellari2004}. The spectrum was first resampled onto a logarithmic wavelength scale without interpolation, but with the masking of bad pixels. Template mismatch was accounted for by simultaneously fitting the continuum of the best-fit \citet{bc2003} template with an additive polynomial, following the same analysis presented in Appendix 3 of \citet{vandesande2013}. 

The effect of template choice can greatly affect the fitted velocity dispersion. We investigated the effect of template choice on the best-fit stellar velocity dispersion, in a similar manner to \citet{vandesande2013}. We fit the spectrum and photometry using FAST with a range of templates for a grid of fixed metallicity and ages. The allowable metallicities were $\mathrm{Z}=0.004$ (super sub-solar), $0.008$ (sub-solar), $0.02$ (solar) and $0.05$ (super-solar).  The age range considered was $\log{\mathrm{Age/yr}}=8.0-9.5$ in increments of 0.1 dex. We increased the resolution of the age grid to 0.05 dex between $\log{\mathrm{Age/yr}}=8.6-8.9$, as we found in previous fitting iterations that the FAST estimated $1\sigma$-error was smaller than 0.1 dex. 

\begin{figure}
  \begin{center}
  \includegraphics[width=\linewidth]{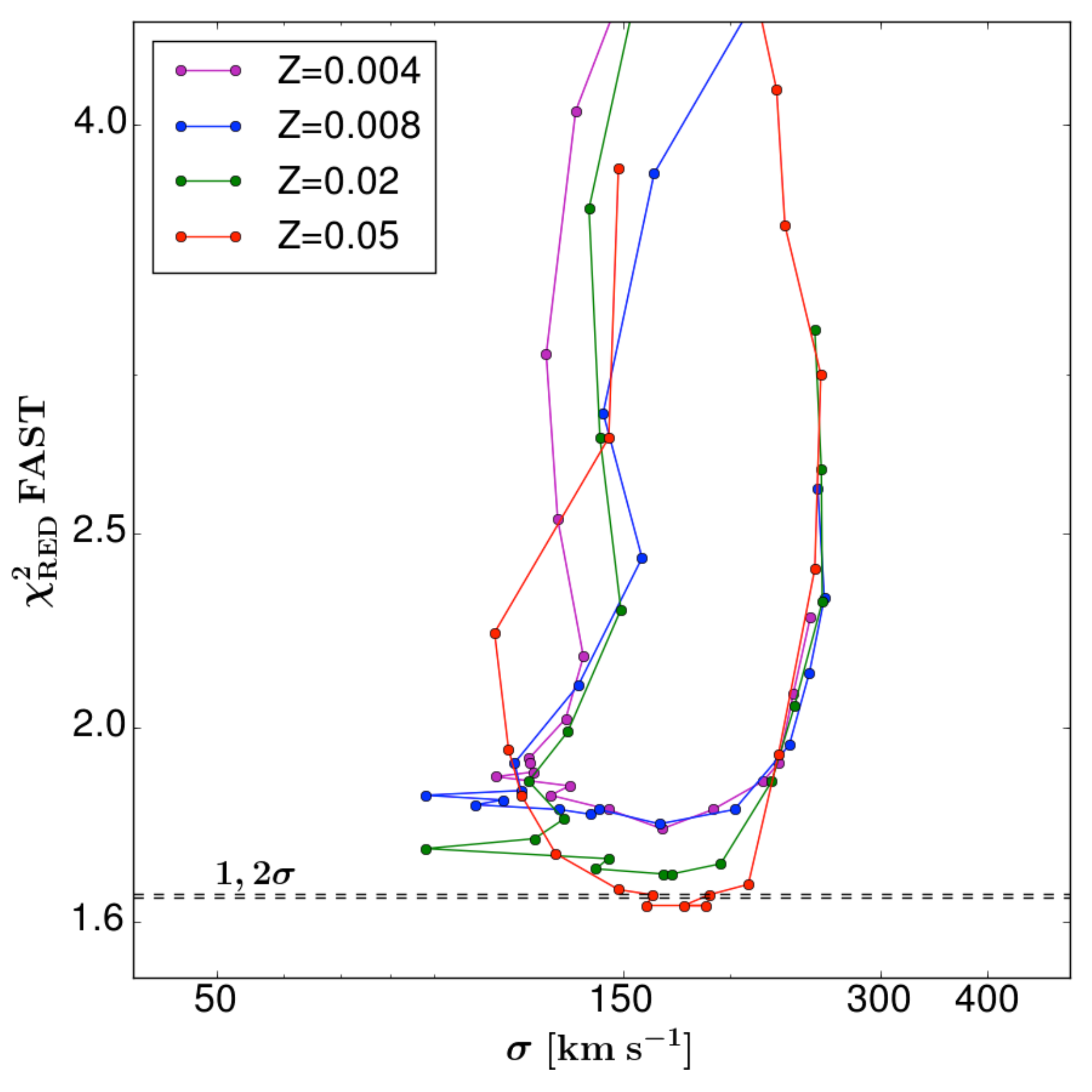}
\caption{The pPXF best-fit stellar velocity dispersion plotted against the $\chi_{\mathrm{red}}^{2}$ output from FAST fitting both the spectrum and the photometry. Each point represents a different age (from $\log{\mathrm{Age/yr}}=8.0-9.5$), and each coloured `track' a different metallicity. The horizontal dashed black lines are the $1\sigma$ and $2\sigma$ upper bounds from the Monte Carlo modelling of the best-fit template at $\log{\mathrm{Age/yr}}=8.75$. Points below these lines indicate templates that are statistically indistinguishable from each other. There are only three of these points, and they result in a stable stellar velocity dispersion implying that template choice is not the dominant source of uncertainty.}
\label{fig:temp}
\end{center}
\end{figure}

In Fig.~\ref{fig:temp}, we show the $\chi_{\mathrm{red}}^{2}$ from FAST as a function of the best fit pPXF stellar velocity dispersion (corrected for template, $\sigma=89~\mathrm{km/s}$, and instrument, $\sigma=25~\mathrm{km/s}$ ,resolution). Using Monte-Carlo simulated errors, we determined 1 and 2 sigma limits on the $\chi_{\mathrm{red}}^{2}$ value of the best-fit FAST model (horizontal black dashed lines in Fig.~\ref{fig:temp}). Points that fall below this line are statistically indistinguishable. We find a very narrow range of statistically indistinguishable templates, concluding that our error will be dominated by the formal errors of the fit, and not the template choice. 

%velocity dispersion dependence on template

Accounting for instrumental and template resolution, as well as a rectangular aperture and seeing (see \citealt{vandesande2013}), we find a best-fit pPXF stellar velocity dispersion of $\sigma = 187\pm43~\mathrm{km~s^{-1}}$ (the error is the $68\%$ confidence limits from the Monte-Carlo simulations). 

From the assumption that the galaxy is virialized, we can determine the dynamical mass as follows:

\begin{equation}
M_{dyn} = \frac{\beta(n)\sigma_{e}^{2}r_{e}}{G}
\label{eq:vir}
\end{equation}

Here G is the gravitational constant,  and $\beta$ is an expression as a function of the Sersic index, $n$, from \citet{cappellari2006} (their equation $20$):

\begin{equation}
\beta(n) = 8.87 - 0.831n + 0.0241n^{2}
\label{eq:beta}
\end{equation}

With $\mathrm{r_{e}}=0.86^{+0.19}_{-0.14}~\mathrm{kpc}$ and $3.50^{+0.68}_{-0.60}$ from Muzzin et al. (in prep), the dynamical mass is $\log{\mathrm{M_{dyn}/\mathrm{M}_{\odot}}}=10.65^{+0.18}_{-0.23}$ which is quite similar to, but slightly above the derived stellar mass of $\log{M_{*}/\mathrm{M_{\odot}}}=10.59^{+0.04}_{-0.05}$. 

\begin{figure*}
  \begin{center}
  \includegraphics[width=\textwidth]{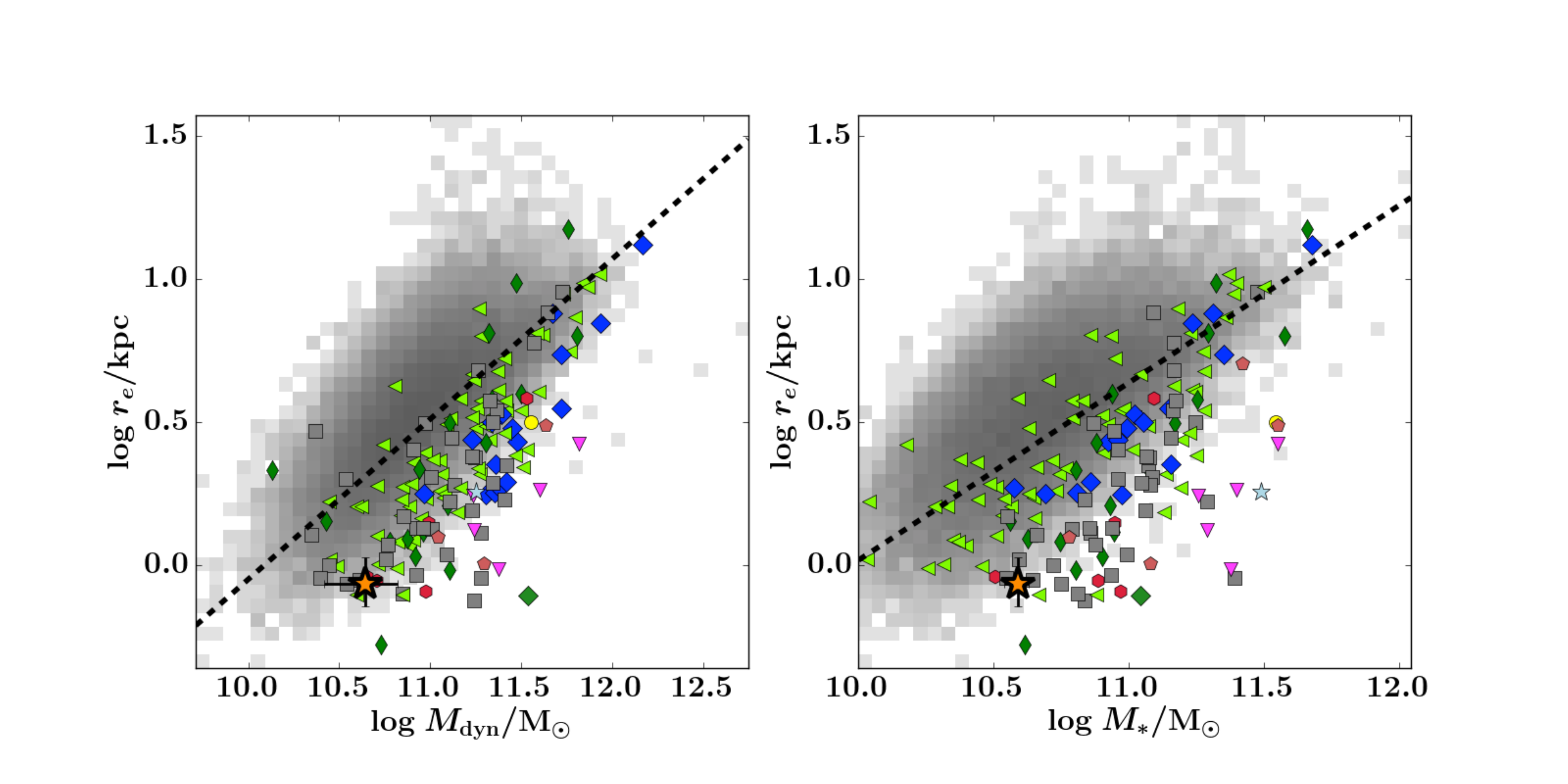}
\caption{Left: The dynamical mass ($\mathrm{M_{dyn}}$) versus $r_{\mathrm{e}}$ of the object of this study (orange star) plotted along side other high redshift objects compiled from \citealt{vandesande2015} where the colored symbols follow the same conventions as the legend in Fig.~\ref{fig:uvj}. The dashed black line is the parameterized mass-size relation of SDSS $z\sim0$ quiescent galaxies, with the best fit parameters from \citealt{vandesande2011}. Right: The same as the left plot, but with stellar mass instead of dynamical mass. In both instances, our object falls below the mass-size relation of $z\sim0$ quiescent galaxies confirming compactness out to lower masses at high redshift.}
\label{fig:mdyn_lmass_re}
\end{center}
\end{figure*}

Fig.~\ref{fig:mdyn_lmass_re} shows $\mathrm{r_{e}}$ as a function of stellar and dynamical mass for high and low-z galaxies with measured stellar velocity dispersions. Also included in Fig.~\ref{fig:mdyn_lmass_re} is the quiescent galaxy mass-size relation for $z\sim0$ (dashed black line). In both stellar and dynamical mass we see that our galaxy is smaller than $z\sim0$ galaxies at equivalent mass, and consistent with the higher-redshift quiescent population. This galaxy is indeed compact, which is well established for quiescent galaxies at these redshifts (see Sec.~\ref{sec:intro} and references therein). 

%mdyn_lmass_re

\begin{figure}
  \begin{center}
  \includegraphics[width=\linewidth]{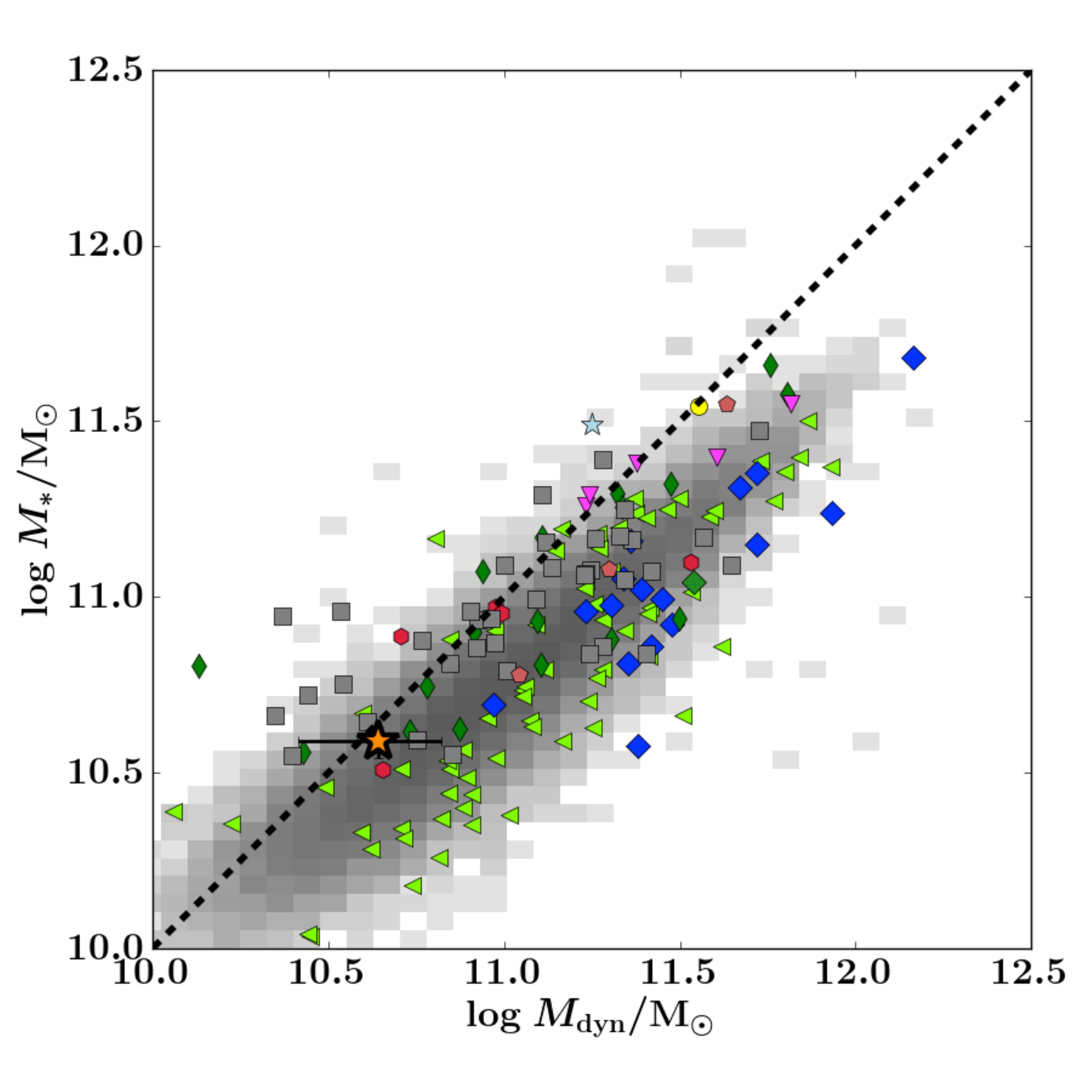}
  \caption{The dynamical ($\mathrm{M_{dyn}}$) versus the stellar ($\mathrm{M}_{*}$) mass of the object of this study (orange star) plotted along side other high-redshift objects compiled from \citealt{vandesande2015} where the colored symbols follow the same conventions as the legend in Fig.~\ref{fig:uvj}. The high-z objects are mass and colour selected. SDSS galaxies are the grey points (which have been colour selected with the same UVJ selection as found in Fig.~\ref{fig:uvj}), and the dashed black line is unity. For our object, the stellar mass and dynamical mass agree well. }
  \label{fig:mdyn_lmass}
  \end{center}
\end{figure}

Fig~\ref{fig:mdyn_lmass} shows the comparison between the FAST determined $M_{*}$, and the $M_{\mathrm{dyn}}$. The black dashed line in Fig.~\ref{fig:mdyn_lmass} is unity - above this line is an `unphysical' regime where the stellar mass exceeds the dynamical mass. From this figure, we see that the implied dark matter fraction appears to be very low, at least in the central regions of the galaxy where the bulk of the light is found (Fig.~\ref{fig:mdyn_lmass_re}). However the error bars are large, and a dark matter fraction of $50\%$ is also consistent with the data, therefore a definitive conclusion can not be drawn. 

%Mdyn_lmass

%Redshift distribution of high redshift galaxies 
\begin{figure}
  \begin{center}
  \includegraphics[width=\linewidth]{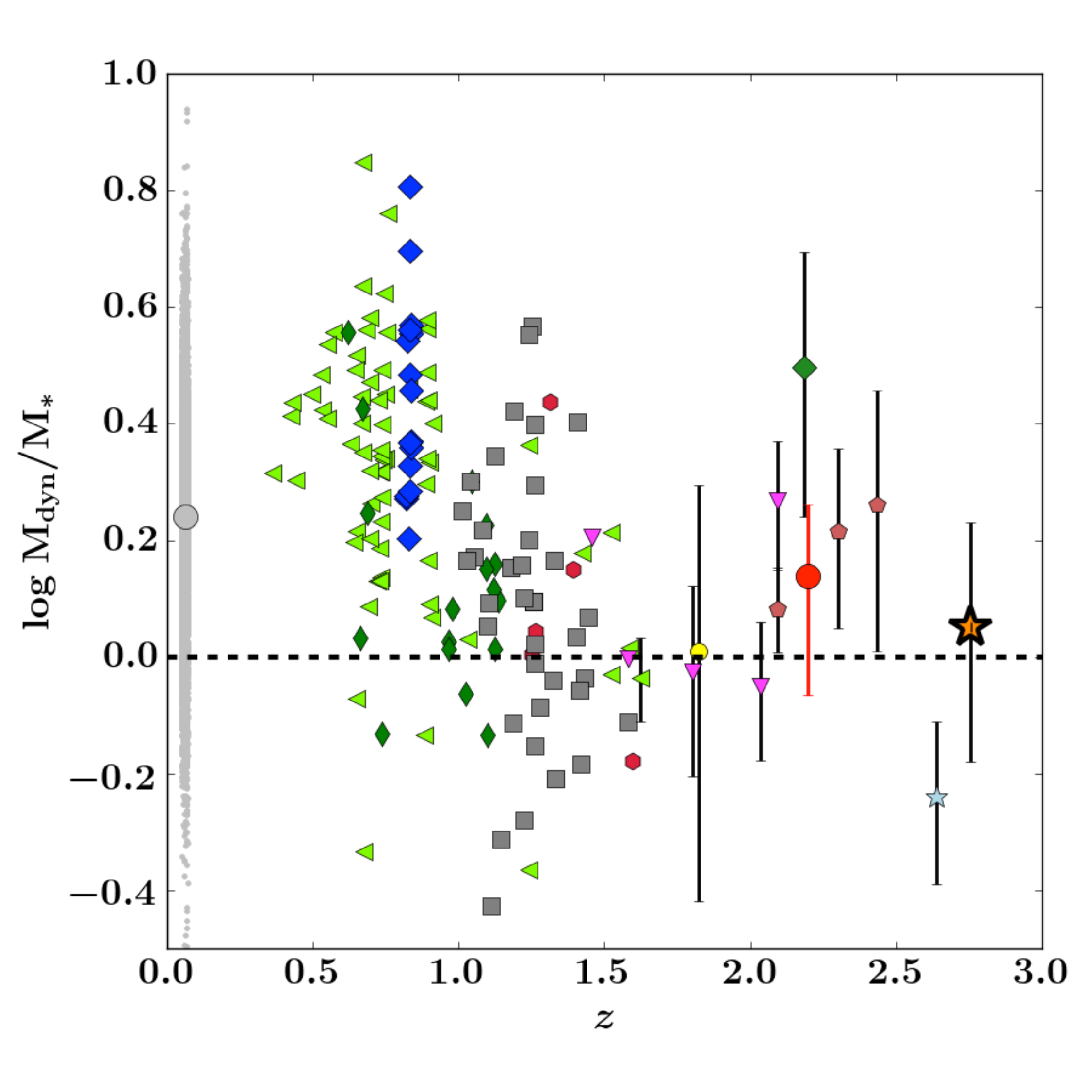}
  \caption{Redshift vs. ratio of dynamical to stellar masses of the color-selected sample compiled by \citet{vandesande2015}, and our object. Grey points are SDSS galaxies following the same color selection.  The colored symbols follow the same conventions as the legend in Fig.~\ref{fig:uvj}. The grey circle is the average total-to-stellar mass ratio for SDSS galaxies. We note that the 68\% confidence limit for the average total-to-stellar mass ratio for the SDSS galaxies is smaller than the grey symbol. The large red circle is the average total-to-stellar mass ratio for galaxies at $z>1.6$ with 68\% confidence limits. Our object is the highest redshift for which a dispersion has been measured.}
  \label{fig:redshift}
  \end{center}
\end{figure}

In Fig~\ref{fig:redshift}, we directly compare the total-to-stellar mass ratio (and thus, the dark matter fraction) as a function of redshift, where the mass fraction approaches unity (also seen in Fig.~\ref{fig:mdyn_lmass}). This figure contains the same literature compilation of high-z quiescent galaxies with velocity dispersion measurements as found in Fig.~\ref{fig:uvj}. Of the known objects with a velocity dispersion, ours is at the highest redshift, and one of the lowest total-to-stellar mass ratios. There is also a weak trend, with quiescent galaxies becoming increasingly baryon dominated with redshift.  This trend was also noted by \citet{vandesande2013,vandesande2015} whose sample extended out to $z=2.3$, however between $z=0$ and $z=1.6$, \citet{belli2014a} found no statistically significant evolution. This may suggest a rapid evolution between redshift $z\sim3$ and $z\sim1.6$. To test this, we compare the average total-to-stellar mass ratio of the SDSS sample to galaxies above redshift $z>1.6$. In Fig.~\ref{fig:redshift}, we have indicated the SDSS median total-to-stellar mass ratio as grey circle. The average for galaxies above $z>1.6$ is indicated by the red symbol with the 68\% confidence interval. From Fig.~\ref{fig:redshift}, we see that galaxies at $z>1.6$ do have a lower total-to-stellar mass ratio consistent with the findings of \citet{vandesande2013,vandesande2015}, however the average value for the SDSS galaxies and those above $z>1.6$ are not statistically different (i.e. they fall within the $2.5\sigma$ uncertainty).

\subsection{Mass Fundamental Plane}
\label{sec:mfp}

Given a stellar mass, a precisely determined $\mathrm{r_{e}}$, and our stellar velocity dispersion measurement, we are able to tentatively explore the MFP to $z\sim3$, as well as lower masses. The most salient difference between the FP and MFP with regards to evolution with redshift is the zero-point. Although the zero-point of the traditional luminosity FP is shown to evolve with redshift \citep[e.g][]{vandokkum1996,vandokkum1998,treu1999,treu2001b}, the MFP does not \citep{bolton2008,holden2010,bezanson2013b}. The evolution of the zero-point of the FP can be used to investigate the luminosity evolution of quiescent galaxies, whereas the zero-point evolution in the MFP can be used to investigate the corresponding structural and dynamical evolution \citep{bezanson2013b}.

Strikingly, \citet{bezanson2013b} established that the zero-point of the MFP does not evolve significantly with redshift, in spite of evolution in the structure and size of quiescent galaxies since $z\sim2$ (see Sec.~\ref{sec:intro} and references therein). One outstanding question is whether this trend holds to higher redshifts and lower mass. With this data, we are able to explore both issues simultaneously. 

\begin{figure}
  \begin{center}
  \includegraphics[width=\linewidth]{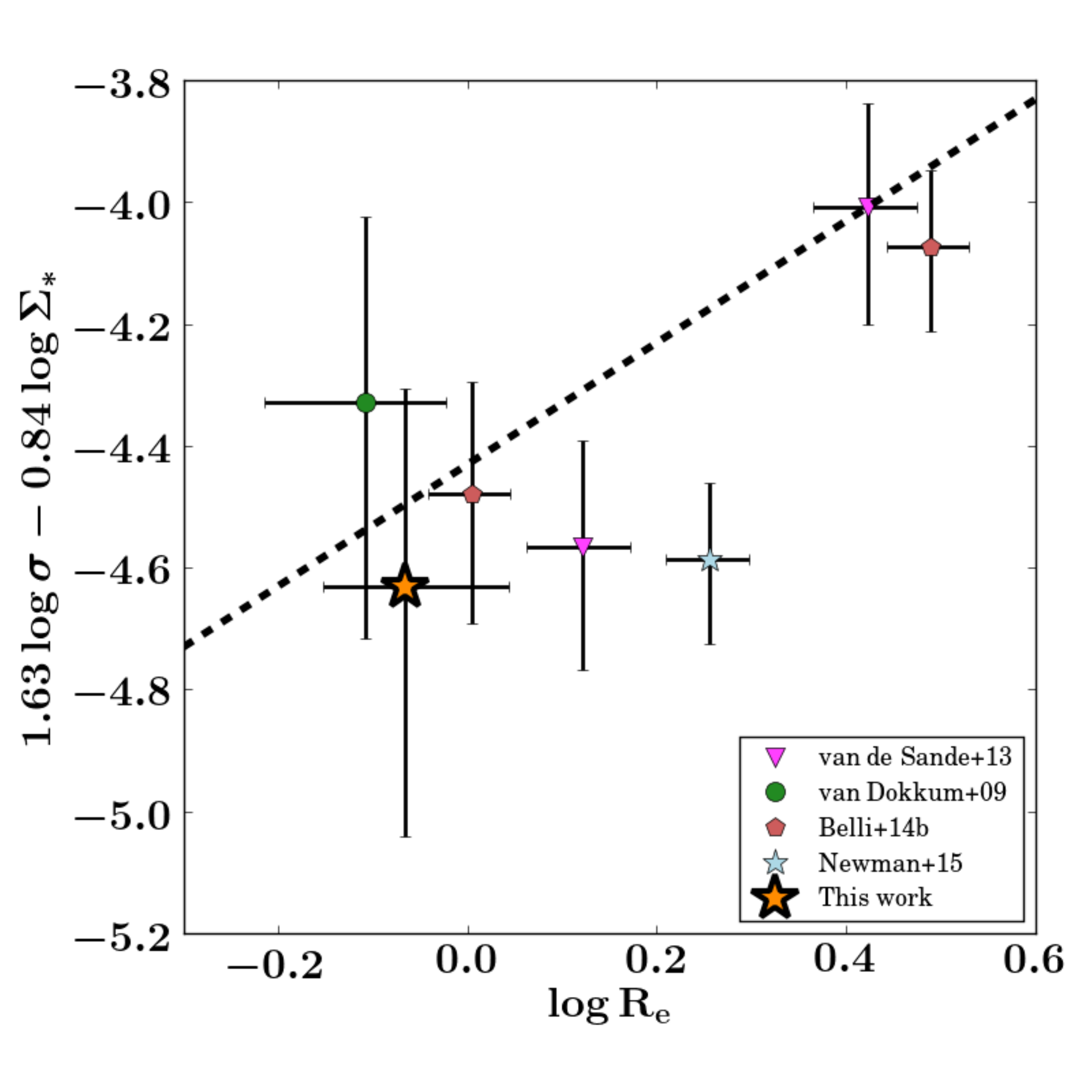}
  \caption{The mass fundamental plane for galaxies at redshift $z>2$. Symbols are as per the previous figures. The black dashed line is the best-fit mass FP between $1.5<z<2.2$ from \citealt{bezanson2013b}. Our galaxy is within the variance of the galaxies at $z>2$ around the mass fundamental plane suggesting it is in place at lower masses.}
  \label{fig:mfp}
  \end{center}
\end{figure}

In Fig.~\ref{fig:mfp} we compare our measurement to galaxies at the highest available redshifts which have velocity dispersions. We applied a redshift cut to the literature compilation of \citet{vandesande2015} of $z>2$, which left only 5 galaxies (where the highest spectroscopic redshift is $z_{\mathrm{spec}}=2.636$) for comparison. Fig.~\ref{fig:mfp} shows that, within measurement uncertainty, our galaxy lies on the same MFP as galaxies at $z\sim2$. This coherence between $z\sim2$ and $z\sim3$ suggests that the MFP evolves very little between these epochs. Further data at these redshifts are required for a definitive conclusion. 

%Mass FP

\section{Discussion and Conclusions}

We have obtained an X-Shooter VLT spectrum of the multiply imaged lensed galaxy COSMOS 0050+4901 found serendipitously in the UltraVISTA field. The lensing of this quiescent galaxy was fortunate, providing a detailed, 'sneak-peak' at the universe during an exciting time. In order to obtain a spectrum with equivalent S/N, without the magnifying effects of gravitational-lensing, 215 hours of integration time on a 10m class telescope would have been required. The existence of only a handful of gravitationally-lensed, quiescent galaxies \citep{auger2011, geier2013, newman2015} highlights the rarity of these objects, and the opportunity they provide to study the high-z, intermediate-mass mass universe preceding the era of JWST. 

At $z_{spec}=2.756\pm0.001$, COSMOS 0050+4901 is one of the highest-redshift quiescent galaxies with a spectroscopic redshift, as well as the highest-redshift galaxy with a measured velocity dispersion ($\sigma = 187\pm43~\mathrm{km~s^{-1}}$). With this spectrum we have detected a suite of Balmer lines ($\mathrm{H}\gamma$, $\mathrm{H}\delta$), and the CaH\&K absorption lines in the observed H-band (Fig.~\ref{fig:zoomin}). The detection of multiple absorption lines provides tight constraints on $z_{spec}$, and is additionally indicative of the presence of older stellar populations. Within the wavelength covered by the spectrum, we find no evidence of emission lines. With our spectrum, the understanding of the stellar populations of this galaxy change from an older ($\log{\mathrm{Age/yr}}=9.0^{+0.2}_{-0.2}$), massive ($\log{\mathrm{M_{*}/M_{\odot}}}=10.82^{+0.05}_{-0.07}$), dusty ($A_{V}=0.9^{+0.2}_{-0.6}$) galaxy (as found by M12) to a younger ($\log{\mathrm{Age/yr}}=8.75^{+0.07}_{-0.07}$), post-starburst galaxy of intermediate mass ($\log{\mathrm{M_{*}/M_{\odot}}}=10.59^{+0.04}_{-0.05}$). 

This new age determination is supported by spectral diagnostics such as the $\mathrm{D_{n}(4000)}$ \citep{balogh1999}, and the Lick index $\mathrm{H}\delta_{A}$ \citep{worthey1997} which have been shown to be sensitive to age \citep{kauffmann2003}. In Fig.~\ref{fig:dn4000_hdelt}, we show how the aforementioned spectral diagnostics reaffirm the younger age determination of this study in a model independent way. Equipped with this best-fit age, the rest-frame optical colors (Fig.~\ref{fig:uvj}), and star formation rate from stellar population modelling (Fig.~\ref{fig:zoomout}) we confirm that intermediate-mass galaxies which halt in-situ star formation are in place by $z\sim3$.

In addition to the brightening effects of the magnification, the magnifying effects also increase spatial resolution. With this increase in resolution, accurate structural parameters are determined. Muzzin et al. (in prep) modelled the gravitationally lensed system and determined the magnification, as well as the surface brightness profile, measuring a sersic index of $3.50^{+0.68}_{-0.60}$ and an $\mathrm{r_{e}}=0.86^{+0.19}_{-0.14}~\mathrm{kpc}$. The right-hand panel of Fig.~\ref{fig:mdyn_lmass_re} shows the precise $\mathrm{r_{e}}$ measurement of Muzzin et al. (in prep) as a function of the stellar mass. Over-plotted for comparison are a local, and high-redshift sample. Like the high-redshift sample, our object falls below the local size-mass relation \citep{shen2003}. We confirm, with a high degree of precision, that this galaxy is compact, which is consistent with what is seen for quiescent galaxies at $z\sim2$ \citep[e.g.,][]{daddi2005, trujillo2006, zirm2007, vandokkum2008, szomoru2012}.

Spectroscopy also allows for a kinematic determination of the mass. The measurement of a dynamical mass is important, as it provides a direct, kinematic method of probing the total matter content of a galaxy, without the uncertainties and prior assumptions associated with parameters such as distance measurement, IMF, and dust content (i.e \citealt{conroy2009} places the uncertainty associated with stellar mass estimates to be $0.6~\mathrm{dex}$ at $z\sim2$). With the dynamical mass we are able to place a strict upper limit on the baryonic contribution to the total mass of the galaxy. In Fig.~\ref{fig:mdyn_lmass} we compare the two measurements, and find the stellar and dynamical masses to be consistent with each other, although our results suggest either a low dark matter fraction in the inner kpc of the galaxy, where the bulk of the light is found, or perhaps that the stellar mass content is overestimated within the parameters discussed by \citet{conroy2009} (such as assumptions about the IMF, dust content). However, Fig.~\ref{fig:dn4000_hdelt} shows, via the $H\delta_{A}$ and $D_{n}4000$ indices, a model independent estimation of the age of this object. This result implies consistency in the model choice. 

With a kinematic determination of the mass, and accurately measured structural parameters ($\mathrm{r_{e}}$, and the Sersic index $n$), we are able to place COSMOS~0050+4901 on the MFP (Fig.~\ref{fig:mfp}). It is well established that local, quiescent galaxies fall on a FP described by surface brightness, size and stellar velocity dispersion \citep{dressler1987,djorgovski1987}, which is tilted with respect to the prediction from virial equilibrium. This tilt does not evolve significantly, but there is an offset in the plane which becomes larger towards higher redshifts \citep[e.g][]{vandokkum1996,vandokkum1998,treu1999,treu2001b} as a result of the luminosity evolution of these galaxies with cosmic time. When replacing the surface brightness with stellar mass density (i.e. the MFP), \citet{bezanson2013b} found that this offset does not evolve significantly and these galaxies fall on the same MFP out to $z\sim2$. In Fig.~\ref{fig:mfp} we compare the object in this study to the highest redshift galaxies available with a velocity dispersion measurement. We show that out to $z\sim3$ quiescent galaxies fall on the same MFP, and that little evolution takes place between $z\sim3$ and the present day.  

A dynamical mass additionally enables the investigation of dark-matter content in the central regions of this galaxy. In Fig.~\ref{fig:redshift}, we have plotted the total-to-stellar mass ratio as a function of redshift, which shows evidence of a trend to decreasing values. However this trend is not statistically significant, as the uncertainties on the dynamical mass at high-$z$ are large. More spectra of these types of objects are required to make a statistically significant claim. If this ratio is low, it is what is expected in the case of a gas-rich, major merger \citep{robertson2006, hopkins2009b}, which implies that this galaxy could represent the first generation of quiescent galaxies in the hierarchical merging scenario \citep{white1978}, making this redshift epoch an exciting prospect for study. 

This case study stands as a proof of concept of the utility of lensed, red galaxies in studying the population of passive galaxies down to lower masses as well as to higher redshifts. This paper also illustrates that few rest-frame optical spectra of quiescent galaxies exist beyond $z>2$, and even fewer of galaxies at intermediate mass. We stress the need for further spectra of passive galaxies at higher-redshift, as well as to lower masses, as this parameter space is still under-explored.

\section{Acknowledgments}

This research has made use of NASA's Astrophysics Data System. The authors also wish to thank the anonymous referee who's suggestions improved the presentation, and overall clarity of this paper.

%Bibliography
\bibliographystyle{apj}
\bibliography{hilletal2015}

\begin{thebibliography}{77}
\expandafter\ifx\csname natexlab\endcsname\relax\def\natexlab#1{#1}\fi

\bibitem[{{Auger} {et~al.}(2011){Auger}, {Treu}, {Brewer}, \&
  {Marshall}}]{auger2011}
{Auger}, M.~W., {Treu}, T., {Brewer}, B.~J., \& {Marshall}, P.~J. 2011, \mnras,
  411, L6

\bibitem[{{Balogh} {et~al.}(1999){Balogh}, {Morris}, {Yee}, {Carlberg}, \&
  {Ellingson}}]{balogh1999}
{Balogh}, M.~L., {Morris}, S.~L., {Yee}, H.~K.~C., {Carlberg}, R.~G., \&
  {Ellingson}, E. 1999, \apj, 527, 54

\bibitem[{{Belli} {et~al.}(2014{\natexlab{a}}){Belli}, {Newman}, \&
  {Ellis}}]{belli2014a}
{Belli}, S., {Newman}, A.~B., \& {Ellis}, R.~S. 2014{\natexlab{a}}, \apj, 783,
  117

\bibitem[{{Belli} {et~al.}(2014{\natexlab{b}}){Belli}, {Newman}, {Ellis}, \&
  {Konidaris}}]{belli2014b}
{Belli}, S., {Newman}, A.~B., {Ellis}, R.~S., \& {Konidaris}, N.~P.
  2014{\natexlab{b}}, \apjl, 788, L29

\bibitem[{{Bernardi} {et~al.}(2003){Bernardi}, {Sheth}, {Annis}, {Burles},
  {Eisenstein}, {Finkbeiner}, {Hogg}, {Lupton}, {Schlegel}, {SubbaRao},
  {Bahcall}, {Blakeslee}, {Brinkmann}, {Castander}, {Connolly}, {Csabai},
  {Doi}, {Fukugita}, {Frieman}, {Heckman}, {Hennessy}, {Ivezi{\'c}}, {Knapp},
  {Lamb}, {McKay}, {Munn}, {Nichol}, {Okamura}, {Schneider}, {Thakar}, \&
  {York}}]{bernardi2003}
{Bernardi}, M., {Sheth}, R.~K., {Annis}, J., {et~al.} 2003, \aj, 125, 1866

\bibitem[{{Bezanson} {et~al.}(2013){Bezanson}, {van Dokkum}, {van de Sande},
  {Franx}, {Leja}, \& {Kriek}}]{bezanson2013b}
{Bezanson}, R., {van Dokkum}, P.~G., {van de Sande}, J., {et~al.} 2013, \apjl,
  779, L21

\bibitem[{{Bolton} {et~al.}(2007){Bolton}, {Burles}, {Treu}, {Koopmans}, \&
  {Moustakas}}]{bolton2007}
{Bolton}, A.~S., {Burles}, S., {Treu}, T., {Koopmans}, L.~V.~E., \&
  {Moustakas}, L.~A. 2007, \apjl, 665, L105

\bibitem[{{Bolton} {et~al.}(2008){Bolton}, {Treu}, {Koopmans}, {Gavazzi},
  {Moustakas}, {Burles}, {Schlegel}, \& {Wayth}}]{bolton2008}
{Bolton}, A.~S., {Treu}, T., {Koopmans}, L.~V.~E., {et~al.} 2008, \apj, 684,
  248

\bibitem[{{Brammer} {et~al.}(2008){Brammer}, {van Dokkum}, \&
  {Coppi}}]{brammer2008}
{Brammer}, G.~B., {van Dokkum}, P.~G., \& {Coppi}, P. 2008, \apj, 686, 1503

\bibitem[{{Brammer} {et~al.}(2012){Brammer}, {S{\'a}nchez-Janssen},
  {Labb{\'e}}, {da Cunha}, {Erb}, {Franx}, {Fumagalli}, {Lundgren},
  {Marchesini}, {Momcheva}, {Nelson}, {Patel}, {Quadri}, {Rix}, {Skelton},
  {Schmidt}, {van der Wel}, {van Dokkum}, {Wake}, \& {Whitaker}}]{brammer2012}
{Brammer}, G.~B., {S{\'a}nchez-Janssen}, R., {Labb{\'e}}, I., {et~al.} 2012,
  \apjl, 758, L17

\bibitem[{{Bruzual} \& {Charlot}(2003)}]{bc2003}
{Bruzual}, G., \& {Charlot}, S. 2003, \mnras, 344, 1000

\bibitem[{{Busarello} {et~al.}(1997){Busarello}, {Capaccioli}, {Capozziello},
  {Longo}, \& {Puddu}}]{busarello1997}
{Busarello}, G., {Capaccioli}, M., {Capozziello}, S., {Longo}, G., \& {Puddu},
  E. 1997, \aap, 320, 415

\bibitem[{{Calzetti} {et~al.}(2000){Calzetti}, {Armus}, {Bohlin}, {Kinney},
  {Koornneef}, \& {Storchi-Bergmann}}]{calzetti2000}
{Calzetti}, D., {Armus}, L., {Bohlin}, R.~C., {et~al.} 2000, \apj, 533, 682

\bibitem[{{Capelato} {et~al.}(1995){Capelato}, {de Carvalho}, \&
  {Carlberg}}]{capelato1995}
{Capelato}, H.~V., {de Carvalho}, R.~R., \& {Carlberg}, R.~G. 1995, \apj, 451,
  525

\bibitem[{{Cappellari} \& {Emsellem}(2004)}]{cappellari2004}
{Cappellari}, M., \& {Emsellem}, E. 2004, \pasp, 116, 138

\bibitem[{{Cappellari} {et~al.}(2006){Cappellari}, {Bacon}, {Bureau}, {Damen},
  {Davies}, {de Zeeuw}, {Emsellem}, {Falc{\'o}n-Barroso}, {Krajnovi{\'c}},
  {Kuntschner}, {McDermid}, {Peletier}, {Sarzi}, {van den Bosch}, \& {van de
  Ven}}]{cappellari2006}
{Cappellari}, M., {Bacon}, R., {Bureau}, M., {et~al.} 2006, \mnras, 366, 1126

\bibitem[{{Cappellari} {et~al.}(2013){Cappellari}, {Scott}, {Alatalo}, {Blitz},
  {Bois}, {Bournaud}, {Bureau}, {Crocker}, {Davies}, {Davis}, {de Zeeuw},
  {Duc}, {Emsellem}, {Khochfar}, {Krajnovi{\'c}}, {Kuntschner}, {McDermid},
  {Morganti}, {Naab}, {Oosterloo}, {Sarzi}, {Serra}, {Weijmans}, \&
  {Young}}]{cappellari2013}
{Cappellari}, M., {Scott}, N., {Alatalo}, K., {et~al.} 2013, \mnras, 432, 1709

\bibitem[{{Chabrier}(2003)}]{chabrier2003}
{Chabrier}, G. 2003, \pasp, 115, 763

\bibitem[{{Conroy} {et~al.}(2009){Conroy}, {Gunn}, \& {White}}]{conroy2009}
{Conroy}, C., {Gunn}, J.~E., \& {White}, M. 2009, \apj, 699, 486

\bibitem[{{Daddi} {et~al.}(2005){Daddi}, {Renzini}, {Pirzkal}, {Cimatti},
  {Malhotra}, {Stiavelli}, {Xu}, {Pasquali}, {Rhoads}, {Brusa}, {di Serego
  Alighieri}, {Ferguson}, {Koekemoer}, {Moustakas}, {Panagia}, \&
  {Windhorst}}]{daddi2005}
{Daddi}, E., {Renzini}, A., {Pirzkal}, N., {et~al.} 2005, \apj, 626, 680

\bibitem[{{Dale} \& {Helou}(2002)}]{dale2002}
{Dale}, D.~A., \& {Helou}, G. 2002, \apj, 576, 159

\bibitem[{{Djorgovski} \& {Davis}(1987)}]{djorgovski1987}
{Djorgovski}, S., \& {Davis}, M. 1987, \apj, 313, 59

\bibitem[{{D'Odorico} {et~al.}(2006){D'Odorico}, {Dekker}, {Mazzoleni},
  {Vernet}, {Guinouard}, {Groot}, {Hammer}, {Rasmussen}, {Kaper}, {Navarro},
  {Pallavicini}, {Peroux}, \& {Zerbi}}]{dodorico2006}
{D'Odorico}, S., {Dekker}, H., {Mazzoleni}, R., {et~al.} 2006, in Society of
  Photo-Optical Instrumentation Engineers (SPIE) Conference Series, Vol. 6269,
  Society of Photo-Optical Instrumentation Engineers (SPIE) Conference Series,
  33

\bibitem[{{Dressler} {et~al.}(1987){Dressler}, {Lynden-Bell}, {Burstein},
  {Davies}, {Faber}, {Terlevich}, \& {Wegner}}]{dressler1987}
{Dressler}, A., {Lynden-Bell}, D., {Burstein}, D., {et~al.} 1987, \apj, 313, 42

\bibitem[{{Franx} {et~al.}(2003){Franx}, {Labb{\'e}}, {Rudnick}, {van Dokkum},
  {Daddi}, {F{\"o}rster Schreiber}, {Moorwood}, {Rix}, {R{\"o}ttgering}, {van
  der Wel}, {van der Werf}, \& {van Starkenburg}}]{franx2003}
{Franx}, M., {Labb{\'e}}, I., {Rudnick}, G., {et~al.} 2003, \apjl, 587, L79

\bibitem[{{Geier} {et~al.}(2013){Geier}, {Richard}, {Man}, {Kr{\"u}hler},
  {Toft}, {Marchesini}, \& {Fynbo}}]{geier2013}
{Geier}, S., {Richard}, J., {Man}, A.~W.~S., {et~al.} 2013, \apj, 777, 87

\bibitem[{{Holden} {et~al.}(2010){Holden}, {van der Wel}, {Kelson}, {Franx}, \&
  {Illingworth}}]{holden2010}
{Holden}, B.~P., {van der Wel}, A., {Kelson}, D.~D., {Franx}, M., \&
  {Illingworth}, G.~D. 2010, \apj, 724, 714

\bibitem[{{Hopkins} {et~al.}(2009){Hopkins}, {Lauer}, {Cox}, {Hernquist}, \&
  {Kormendy}}]{hopkins2009b}
{Hopkins}, P.~F., {Lauer}, T.~R., {Cox}, T.~J., {Hernquist}, L., \& {Kormendy},
  J. 2009, \apjs, 181, 486

\bibitem[{{Ilbert} {et~al.}(2013){Ilbert}, {McCracken}, {Le F{\`e}vre},
  {Capak}, {Dunlop}, {Karim}, {Renzini}, {Caputi}, {Boissier}, {Arnouts},
  {Aussel}, {Comparat}, {Guo}, {Hudelot}, {Kartaltepe}, {Kneib}, {Krogager},
  {Le Floc'h}, {Lilly}, {Mellier}, {Milvang-Jensen}, {Moutard}, {Onodera},
  {Richard}, {Salvato}, {Sanders}, {Scoville}, {Silverman}, {Taniguchi},
  {Tasca}, {Thomas}, {Toft}, {Tresse}, {Vergani}, {Wolk}, \&
  {Zirm}}]{ilbert2013}
{Ilbert}, O., {McCracken}, H.~J., {Le F{\`e}vre}, O., {et~al.} 2013, \aap, 556,
  A55

\bibitem[{{Kauffmann} {et~al.}(2003){Kauffmann}, {Heckman}, {White}, {Charlot},
  {Tremonti}, {Brinchmann}, {Bruzual}, {Peng}, {Seibert}, {Bernardi},
  {Blanton}, {Brinkmann}, {Castander}, {Cs{\'a}bai}, {Fukugita}, {Ivezic},
  {Munn}, {Nichol}, {Padmanabhan}, {Thakar}, {Weinberg}, \&
  {York}}]{kauffmann2003}
{Kauffmann}, G., {Heckman}, T.~M., {White}, S.~D.~M., {et~al.} 2003, \mnras,
  341, 33

\bibitem[{{Kriek} {et~al.}(2009){Kriek}, {van Dokkum}, {Labb{\'e}}, {Franx},
  {Illingworth}, {Marchesini}, \& {Quadri}}]{kriek2009}
{Kriek}, M., {van Dokkum}, P.~G., {Labb{\'e}}, I., {et~al.} 2009, \apj, 700,
  221

\bibitem[{{Kriek} {et~al.}(2006){Kriek}, {van Dokkum}, {Franx}, {F{\"o}rster
  Schreiber}, {Gawiser}, {Illingworth}, {Labb{\'e}}, {Marchesini}, {Quadri},
  {Rix}, {Rudnick}, {Toft}, {van der Werf}, \& {Wuyts}}]{kriek2006}
{Kriek}, M., {van Dokkum}, P.~G., {Franx}, M., {et~al.} 2006, \apj, 645, 44

\bibitem[{{Labb{\'e}} {et~al.}(2005){Labb{\'e}}, {Huang}, {Franx}, {Rudnick},
  {Barmby}, {Daddi}, {van Dokkum}, {Fazio}, {Schreiber}, {Moorwood}, {Rix},
  {R{\"o}ttgering}, {Trujillo}, \& {van der Werf}}]{labbe2005}
{Labb{\'e}}, I., {Huang}, J., {Franx}, M., {et~al.} 2005, \apjl, 624, L81

\bibitem[{{McCracken} {et~al.}(2012){McCracken}, {Milvang-Jensen}, {Dunlop},
  {Franx}, {Fynbo}, {Le F{\`e}vre}, {Holt}, {Caputi}, {Goranova}, {Buitrago},
  {Emerson}, {Freudling}, {Hudelot}, {L{\'o}pez-Sanjuan}, {Magnard}, {Mellier},
  {M{\o}ller}, {Nilsson}, {Sutherland}, {Tasca}, \& {Zabl}}]{mccracken2012}
{McCracken}, H.~J., {Milvang-Jensen}, B., {Dunlop}, J., {et~al.} 2012, \aap,
  544, A156

\bibitem[{{Modigliani} {et~al.}(2010){Modigliani}, {Goldoni}, {Royer},
  {Haigron}, {Guglielmi}, {Fran{\c c}ois}, {Horrobin}, {Bristow}, {Vernet},
  {Moehler}, {Kerber}, {Ballester}, {Mason}, \& {Christensen}}]{modigliani2010}
{Modigliani}, A., {Goldoni}, P., {Royer}, F., {et~al.} 2010, in Society of
  Photo-Optical Instrumentation Engineers (SPIE) Conference Series, Vol. 7737,
  Society of Photo-Optical Instrumentation Engineers (SPIE) Conference Series,
  28

\bibitem[{{Muzzin} {et~al.}(2012){Muzzin}, {Labb{\'e}}, {Franx}, {van Dokkum},
  {Holt}, {Szomoru}, {van de Sande}, {Brammer}, {Marchesini}, {Stefanon},
  {Buitrago}, {Caputi}, {Dunlop}, {Fynbo}, {Le F{\'e}vre}, {McCracken}, \&
  {Milvang-Jensen}}]{muzzin2012}
{Muzzin}, A., {Labb{\'e}}, I., {Franx}, M., {et~al.} 2012, \apj, 761, 142

\bibitem[{{Muzzin} {et~al.}(2013{\natexlab{a}}){Muzzin}, {Marchesini},
  {Stefanon}, {Franx}, {Milvang-Jensen}, {Dunlop}, {Fynbo}, {Brammer},
  {Labb{\'e}}, \& {van Dokkum}}]{muzzin2013a}
{Muzzin}, A., {Marchesini}, D., {Stefanon}, M., {et~al.} 2013{\natexlab{a}},
  \apjs, 206, 8

\bibitem[{{Muzzin} {et~al.}(2013{\natexlab{b}}){Muzzin}, {Marchesini},
  {Stefanon}, {Franx}, {McCracken}, {Milvang-Jensen}, {Dunlop}, {Fynbo},
  {Brammer}, {Labb{\'e}}, \& {van Dokkum}}]{muzzin2013b}
---. 2013{\natexlab{b}}, \apj, 777, 18

\bibitem[{{Newman} {et~al.}(2015){Newman}, {Belli}, \& {Ellis}}]{newman2015}
{Newman}, A.~B., {Belli}, S., \& {Ellis}, R.~S. 2015, ArXiv e-prints

\bibitem[{{Newman} {et~al.}(2010){Newman}, {Ellis}, {Treu}, \&
  {Bundy}}]{newman2010}
{Newman}, A.~B., {Ellis}, R.~S., {Treu}, T., \& {Bundy}, K. 2010, \apjl, 717,
  L103

\bibitem[{{Onodera} {et~al.}(2010){Onodera}, {Arimoto}, {Daddi}, {Renzini},
  {Kong}, {Cimatti}, {Broadhurst}, \& {Alexander}}]{onodera2010}
{Onodera}, M., {Arimoto}, N., {Daddi}, E., {et~al.} 2010, \apj, 715, 385

\bibitem[{{Onodera} {et~al.}(2012){Onodera}, {Renzini}, {Carollo},
  {Cappellari}, {Mancini}, {Strazzullo}, {Daddi}, {Arimoto}, {Gobat}, {Yamada},
  {McCracken}, {Ilbert}, {Capak}, {Cimatti}, {Giavalisco}, {Koekemoer}, {Kong},
  {Lilly}, {Motohara}, {Ohta}, {Sanders}, {Scoville}, {Tamura}, \&
  {Taniguchi}}]{onodera2012}
{Onodera}, M., {Renzini}, A., {Carollo}, M., {et~al.} 2012, \apj, 755, 26

\bibitem[{{Pahre} {et~al.}(1995){Pahre}, {Djorgovski}, \& {de
  Carvalho}}]{pahre1995}
{Pahre}, M.~A., {Djorgovski}, S.~G., \& {de Carvalho}, R.~R. 1995, \apjl, 453,
  L17

\bibitem[{{Peng} {et~al.}(2010){Peng}, {Ho}, {Impey}, \& {Rix}}]{peng2010}
{Peng}, C.~Y., {Ho}, L.~C., {Impey}, C.~D., \& {Rix}, H.-W. 2010, \aj, 139,
  2097

\bibitem[{{Robertson} {et~al.}(2006){Robertson}, {Cox}, {Hernquist}, {Franx},
  {Hopkins}, {Martini}, \& {Springel}}]{robertson2006}
{Robertson}, B., {Cox}, T.~J., {Hernquist}, L., {et~al.} 2006, \apj, 641, 21

\bibitem[{{Schreiber} {et~al.}(2015){Schreiber}, {Pannella}, {Elbaz},
  {B{\'e}thermin}, {Inami}, {Dickinson}, {Magnelli}, {Wang}, {Aussel}, {Daddi},
  {Juneau}, {Shu}, {Sargent}, {Buat}, {Faber}, {Ferguson}, {Giavalisco},
  {Koekemoer}, {Magdis}, {Morrison}, {Papovich}, {Santini}, \&
  {Scott}}]{schreiber2015}
{Schreiber}, C., {Pannella}, M., {Elbaz}, D., {et~al.} 2015, \aap, 575, A74

\bibitem[{{Sharon} {et~al.}(2012){Sharon}, {Gladders}, {Rigby}, {Wuyts},
  {Koester}, {Bayliss}, \& {Barrientos}}]{sharon2012}
{Sharon}, K., {Gladders}, M.~D., {Rigby}, J.~R., {et~al.} 2012, \apj, 746, 161

\bibitem[{{Shen} {et~al.}(2003){Shen}, {Mo}, {White}, {Blanton}, {Kauffmann},
  {Voges}, {Brinkmann}, \& {Csabai}}]{shen2003}
{Shen}, S., {Mo}, H.~J., {White}, S.~D.~M., {et~al.} 2003, \mnras, 343, 978

\bibitem[{{Smail} {et~al.}(2007){Smail}, {Swinbank}, {Richard}, {Ebeling},
  {Kneib}, {Edge}, {Stark}, {Ellis}, {Dye}, {Smith}, \& {Mullis}}]{smail2007}
{Smail}, I., {Swinbank}, A.~M., {Richard}, J., {et~al.} 2007, \apjl, 654, L33

\bibitem[{{Stark} {et~al.}(2013){Stark}, {Auger}, {Belokurov}, {Jones},
  {Robertson}, {Ellis}, {Sand}, {Moiseev}, {Eagle}, \& {Myers}}]{stark2013}
{Stark}, D.~P., {Auger}, M., {Belokurov}, V., {et~al.} 2013, \mnras, 436, 1040

\bibitem[{{Straatman} {et~al.}(2014){Straatman}, {Labb{\'e}}, {Spitler},
  {Allen}, {Altieri}, {Brammer}, {Dickinson}, {van Dokkum}, {Inami},
  {Glazebrook}, {Kacprzak}, {Kawinwanichakij}, {Kelson}, {McCarthy},
  {Mehrtens}, {Monson}, {Murphy}, {Papovich}, {Persson}, {Quadri}, {Rees},
  {Tomczak}, {Tran}, \& {Tilvi}}]{straatman2014}
{Straatman}, C.~M.~S., {Labb{\'e}}, I., {Spitler}, L.~R., {et~al.} 2014, \apjl,
  783, L14

\bibitem[{{Szomoru} {et~al.}(2012){Szomoru}, {Franx}, \& {van
  Dokkum}}]{szomoru2012}
{Szomoru}, D., {Franx}, M., \& {van Dokkum}, P.~G. 2012, \apj, 749, 121

\bibitem[{{Toft} {et~al.}(2012){Toft}, {Gallazzi}, {Zirm}, {Wold}, {Zibetti},
  {Grillo}, \& {Man}}]{toft2012}
{Toft}, S., {Gallazzi}, A., {Zirm}, A., {et~al.} 2012, \apj, 754, 3

\bibitem[{{Treu} {et~al.}(2001){Treu}, {Stiavelli}, {Bertin}, {Casertano}, \&
  {M{\o}ller}}]{treu2001b}
{Treu}, T., {Stiavelli}, M., {Bertin}, G., {Casertano}, S., \& {M{\o}ller}, P.
  2001, \mnras, 326, 237

\bibitem[{{Treu} {et~al.}(1999){Treu}, {Stiavelli}, {Casertano}, {M{\o}ller},
  \& {Bertin}}]{treu1999}
{Treu}, T., {Stiavelli}, M., {Casertano}, S., {M{\o}ller}, P., \& {Bertin}, G.
  1999, \mnras, 308, 1037

\bibitem[{{Trujillo} {et~al.}(2006){Trujillo}, {F{\"o}rster Schreiber},
  {Rudnick}, {Barden}, {Franx}, {Rix}, {Caldwell}, {McIntosh}, {Toft},
  {H{\"a}ussler}, {Zirm}, {van Dokkum}, {Labb{\'e}}, {Moorwood},
  {R{\"o}ttgering}, {van der Wel}, {van der Werf}, \& {van
  Starkenburg}}]{trujillo2006}
{Trujillo}, I., {F{\"o}rster Schreiber}, N.~M., {Rudnick}, G., {et~al.} 2006,
  \apj, 650, 18

\bibitem[{{van de Sande} {et~al.}(2014){van de Sande}, {Kriek}, {Franx},
  {Bezanson}, \& {van Dokkum}}]{vandesande2014}
{van de Sande}, J., {Kriek}, M., {Franx}, M., {Bezanson}, R., \& {van Dokkum},
  P.~G. 2014, \apjl, 793, L31

\bibitem[{{van de Sande} {et~al.}(2015){van de Sande}, {Kriek}, {Franx},
  {Bezanson}, \& {van Dokkum}}]{vandesande2015}
---. 2015, \apj, 799, 125

\bibitem[{{van de Sande} {et~al.}(2011){van de Sande}, {Kriek}, {Franx}, {van
  Dokkum}, {Bezanson}, {Whitaker}, {Brammer}, {Labb{\'e}}, {Groot}, \&
  {Kaper}}]{vandesande2011}
{van de Sande}, J., {Kriek}, M., {Franx}, M., {et~al.} 2011, \apjl, 736, L9

\bibitem[{{van de Sande} {et~al.}(2013){van de Sande}, {Kriek}, {Franx}, {van
  Dokkum}, {Bezanson}, {Bouwens}, {Quadri}, {Rix}, \&
  {Skelton}}]{vandesande2013}
---. 2013, \apj, 771, 85

\bibitem[{{van der Wel} {et~al.}(2005){van der Wel}, {Franx}, {van Dokkum},
  {Rix}, {Illingworth}, \& {Rosati}}]{vanderwel2005}
{van der Wel}, A., {Franx}, M., {van Dokkum}, P.~G., {et~al.} 2005, \apj, 631,
  145

\bibitem[{{van der Wel} {et~al.}(2013){van der Wel}, {van de Ven}, {Maseda},
  {Rix}, {Rudnick}, {Grazian}, {Finkelstein}, {Koo}, {Faber}, {Ferguson},
  {Koekemoer}, {Grogin}, \& {Kocevski}}]{vanderwel2013}
{van der Wel}, A., {van de Ven}, G., {Maseda}, M., {et~al.} 2013, \apjl, 777,
  L17

\bibitem[{{van Dokkum} \& {Franx}(1996)}]{vandokkum1996}
{van Dokkum}, P.~G., \& {Franx}, M. 1996, \mnras, 281, 985

\bibitem[{{van Dokkum} {et~al.}(1998){van Dokkum}, {Franx}, {Kelson}, \&
  {Illingworth}}]{vandokkum1998}
{van Dokkum}, P.~G., {Franx}, M., {Kelson}, D.~D., \& {Illingworth}, G.~D.
  1998, \apjl, 504, L17

\bibitem[{{van Dokkum} {et~al.}(2009){van Dokkum}, {Kriek}, \&
  {Franx}}]{vandokkum2009}
{van Dokkum}, P.~G., {Kriek}, M., \& {Franx}, M. 2009, \nat, 460, 717

\bibitem[{{van Dokkum} {et~al.}(2003){van Dokkum}, {F{\"o}rster Schreiber},
  {Franx}, {Daddi}, {Illingworth}, {Labb{\'e}}, {Moorwood}, {Rix},
  {R{\"o}ttgering}, {Rudnick}, {van der Wel}, {van der Werf}, \& {van
  Starkenburg}}]{vandokkum2003}
{van Dokkum}, P.~G., {F{\"o}rster Schreiber}, N.~M., {Franx}, M., {et~al.}
  2003, \apjl, 587, L83

\bibitem[{{van Dokkum} {et~al.}(2006){van Dokkum}, {Quadri}, {Marchesini},
  {Rudnick}, {Franx}, {Gawiser}, {Herrera}, {Wuyts}, {Lira}, {Labb{\'e}},
  {Maza}, {Illingworth}, {F{\"o}rster Schreiber}, {Kriek}, {Rix}, {Taylor},
  {Toft}, {Webb}, \& {Yi}}]{vandokkum2006}
{van Dokkum}, P.~G., {Quadri}, R., {Marchesini}, D., {et~al.} 2006, \apjl, 638,
  L59

\bibitem[{{van Dokkum} {et~al.}(2008){van Dokkum}, {Franx}, {Kriek}, {Holden},
  {Illingworth}, {Magee}, {Bouwens}, {Marchesini}, {Quadri}, {Rudnick},
  {Taylor}, \& {Toft}}]{vandokkum2008}
{van Dokkum}, P.~G., {Franx}, M., {Kriek}, M., {et~al.} 2008, \apjl, 677, L5

\bibitem[{{van Dokkum} {et~al.}(2010){van Dokkum}, {Whitaker}, {Brammer},
  {Franx}, {Kriek}, {Labb{\'e}}, {Marchesini}, {Quadri}, {Bezanson},
  {Illingworth}, {Muzzin}, {Rudnick}, {Tal}, \& {Wake}}]{vandokkum2010}
{van Dokkum}, P.~G., {Whitaker}, K.~E., {Brammer}, G., {et~al.} 2010, \apj,
  709, 1018

\bibitem[{{Vernet} {et~al.}(2011){Vernet}, {Dekker}, {D'Odorico}, {Kaper},
  {Kjaergaard}, {Hammer}, {Randich}, {Zerbi}, {Groot}, {Hjorth}, {Guinouard},
  {Navarro}, {Adolfse}, {Albers}, {Amans}, {Andersen}, {Andersen}, {Binetruy},
  {Bristow}, {Castillo}, {Chemla}, {Christensen}, {Conconi}, {Conzelmann},
  {Dam}, {de Caprio}, {de Ugarte Postigo}, {Delabre}, {di Marcantonio},
  {Downing}, {Elswijk}, {Finger}, {Fischer}, {Flores}, {Fran{\c c}ois},
  {Goldoni}, {Guglielmi}, {Haigron}, {Hanenburg}, {Hendriks}, {Horrobin},
  {Horville}, {Jessen}, {Kerber}, {Kern}, {Kiekebusch}, {Kleszcz}, {Klougart},
  {Kragt}, {Larsen}, {Lizon}, {Lucuix}, {Mainieri}, {Manuputy}, {Martayan},
  {Mason}, {Mazzoleni}, {Michaelsen}, {Modigliani}, {Moehler}, {M{\o}ller},
  {Norup S{\o}rensen}, {N{\o}rregaard}, {P{\'e}roux}, {Patat}, {Pena}, {Pragt},
  {Reinero}, {Rigal}, {Riva}, {Roelfsema}, {Royer}, {Sacco}, {Santin},
  {Schoenmaker}, {Spano}, {Sweers}, {Ter Horst}, {Tintori}, {Tromp}, {van
  Dael}, {van der Vliet}, {Venema}, {Vidali}, {Vinther}, {Vola}, {Winters},
  {Wistisen}, {Wulterkens}, \& {Zacchei}}]{vernet2011}
{Vernet}, J., {Dekker}, H., {D'Odorico}, S., {et~al.} 2011, \aap, 536, A105

\bibitem[{{Vieira} {et~al.}(2013){Vieira}, {Marrone}, {Chapman}, {De Breuck},
  {Hezaveh}, {Wei{$\beta$}}, {Aguirre}, {Aird}, {Aravena}, {Ashby}, {Bayliss},
  {Benson}, {Biggs}, {Bleem}, {Bock}, {Bothwell}, {Bradford}, {Brodwin},
  {Carlstrom}, {Chang}, {Crawford}, {Crites}, {de Haan}, {Dobbs}, {Fomalont},
  {Fassnacht}, {George}, {Gladders}, {Gonzalez}, {Greve}, {Gullberg},
  {Halverson}, {High}, {Holder}, {Holzapfel}, {Hoover}, {Hrubes}, {Hunter},
  {Keisler}, {Lee}, {Leitch}, {Lueker}, {Luong-van}, {Malkan}, {McIntyre},
  {McMahon}, {Mehl}, {Menten}, {Meyer}, {Mocanu}, {Murphy}, {Natoli}, {Padin},
  {Plagge}, {Reichardt}, {Rest}, {Ruel}, {Ruhl}, {Sharon}, {Schaffer}, {Shaw},
  {Shirokoff}, {Spilker}, {Stalder}, {Staniszewski}, {Stark}, {Story},
  {Vanderlinde}, {Welikala}, \& {Williamson}}]{vieira2013}
{Vieira}, J.~D., {Marrone}, D.~P., {Chapman}, S.~C., {et~al.} 2013, \nat, 495,
  344

\bibitem[{{Whitaker} {et~al.}(2011){Whitaker}, {Labb{\'e}}, {van Dokkum},
  {Brammer}, {Kriek}, {Marchesini}, {Quadri}, {Franx}, {Muzzin}, {Williams},
  {Bezanson}, {Illingworth}, {Lee}, {Lundgren}, {Nelson}, {Rudnick}, {Tal}, \&
  {Wake}}]{whitaker2011}
{Whitaker}, K.~E., {Labb{\'e}}, I., {van Dokkum}, P.~G., {et~al.} 2011, \apj,
  735, 86

\bibitem[{{White} \& {Rees}(1978)}]{white1978}
{White}, S.~D.~M., \& {Rees}, M.~J. 1978, \mnras, 183, 341

\bibitem[{{Williams} {et~al.}(2009){Williams}, {Quadri}, {Franx}, {van Dokkum},
  \& {Labb{\'e}}}]{williams2009}
{Williams}, R.~J., {Quadri}, R.~F., {Franx}, M., {van Dokkum}, P., \&
  {Labb{\'e}}, I. 2009, \apj, 691, 1879

\bibitem[{{Worthey} \& {Ottaviani}(1997)}]{worthey1997}
{Worthey}, G., \& {Ottaviani}, D.~L. 1997, \apjs, 111, 377

\bibitem[{{Wuyts} {et~al.}(2004){Wuyts}, {van Dokkum}, {Kelson}, {Franx}, \&
  {Illingworth}}]{wuyts2004}
{Wuyts}, S., {van Dokkum}, P.~G., {Kelson}, D.~D., {Franx}, M., \&
  {Illingworth}, G.~D. 2004, \apj, 605, 677

\bibitem[{{Zirm} {et~al.}(2007){Zirm}, {van der Wel}, {Franx}, {Labb{\'e}},
  {Trujillo}, {van Dokkum}, {Toft}, {Daddi}, {Rudnick}, {Rix},
  {R{\"o}ttgering}, \& {van der Werf}}]{zirm2007}
{Zirm}, A.~W., {van der Wel}, A., {Franx}, M., {et~al.} 2007, \apj, 656, 66

\end{thebibliography}

\end{document}